\documentclass[usenatbib]{mnras}
\usepackage{mathptmx}
\usepackage{color}
\usepackage{xcolor}
\usepackage{graphics,graphicx,amssymb,amsmath}\usepackage{multirow}
\usepackage{soul}
\soulregister\citep7
\soulregister\citet7
\soulregister\ref7

\def\gtrsim{\lower 2pt \hbox{$\, \buildrel {\scriptstyle >}\over
{\scriptstyle \sim}\,$}}
\def\lesssim{\lower 2pt \hbox{$\, \buildrel {\scriptstyle <}\over
{\scriptstyle \sim}\,$}}

\def\hst{{\sl HST}}
\def\XMM{{\sl XMM-Newton}}
\def\xmm{{\sl XMM-Newton}}

\def\chandra{{\sl Chandra}}

\def\HST{{\sl HST}}
\def\hst{{\sl HST}}
\def\wise{{\sl WISE}}
\def\2mass{{\sl 2MASS}}
\def\galex{{\sl GALEX}}

\def\spitzer{{\sl Spitzer}}

\def\micron{{$~\mu$m}}
\def\as{{$^{\prime\prime}$}}
\def\ha{{H$\alpha$}}
\def\hii{H{\small II}}
\def\ovii{O{\small VII}}
\def\oviii{O{\small VIII}}

\newcommand{\lumcgs}{erg~s$^{-1}$}

\def\sou{M83}

\date{Accepted 2021 October 12. Received 2021 September 30; in original form 2021 August 12}
\pubyear{2021}
\begin{document}

\title{Deep \chandra\ observations of diffuse hot plasma in M83}
\author[Q. Daniel Wang et al.]{Q. Daniel Wang$^{1}$\thanks{Contact e-mail:wqd@umass.edu},  
 Yuxuan Zeng$^{1,2,3,4}$, \'{A}kos Bogd\'{a}n$^{5}$, \& Li Ji$^{2,3}$\\
$^{1}$ Department of Astronomy, University of Massachusetts,  Amherst, MA 01003, USA \\
$^{2}$ Purple Mountain Observatory, Chinese Academy of Sciences, People's Republic of China \\
$^{3}$ Key Laboratory of Dark Matter and Space Astronomy, CAS, People's Republic of China\\
$^{4}$ Kapteyn Astronomical Institute, University of Groningen, PO Box 800, 9700 AV Groningen, The Netherlands \\
$^{5}$ Harvard-Smithsonian Center for Astrophysics, 60 Garden Street, Cambridge, MA 02138, USA\\
}
\maketitle
\begin{abstract}
It is widely believed that galaxy formation and evolution is regulated by stellar mechanical feedback in forms of fast stellar winds and supernova explosions. However, the coupling of this feedback with the interstellar medium remains poorly understood. We examine how the coupling may be traced by diffuse soft X-ray emission in M83 -- a nearby face-on spiral galaxy undergoing active star formation, based chiefly on 729~ks Chandra observations.  Our main findings are 1) the X-ray emission is enhanced not only along the galaxy's grand spiral arms, but also clearly in their downstreams; 2)  the spectrum of the emission can be well characterized by a super-solar metallicity plasma with a lognormal temperature distribution, plus an X-ray absorption of a lognormal column density distribution; 3) the intensity of the emission is strongly anti-correlated with the dust obscuration seen in optical images of the galaxy.  These findings suggest A) the morphology of the X-ray emission is likely due to the convolution of the feedback heating of the plasma with its thermal and dynamical evolution; B) the X-ray emission, accounting for $\sim 10\%$ of the feedback energy input rate, probably traces only the high-energy tail of the radiation from the plasma; C) a good fraction of the recent star forming regions seems sufficiently energetic to produce multi-phased outflows, likely responsible for much of the dust obscuration and X-ray absorption. Direct confrontation of the findings with theories/simulations could help to understand the underlying astrophysics of the coupling and how the hot plasma shapes the interstellar medium.
\end{abstract}

\section{Introduction}\label{s:int}

Stellar feedback is an important part of our present understanding of galaxy formation and evolution \citep[e.g.,][]{Hopkins2014,Fujimoto2016,Kelly2020}.  However, much of this understanding is still not based on first principles, rather on recipes for subgrid astrophysics, which have been subject to few direct observational tests, leaving us with great uncertainties as to how the feedback operates and which prescription of physics might be closest to reality.
 \citep[e.g.,][]{Somerville2015,Li2020,Hopkins2021,Quataert2021,Afruni2021,Jeffreson2021}. We do have a solid framework for the feedback from high-mass stars in the form of fast stellar winds, core-collapsed supernovae (CC-SNe) and radiation, although its coupling with the surrounding interstellar medium (ISM)  is poorly understood.  The coupling cannot be well simulated, because of its multiphase nature and enormous scale range from individual stars to galaxies and beyond to the intergalactic medium. Therefore, the true impact of the stellar feedback on galaxy formation and evolution remains uncertain.

Diffuse X-ray emission has commonly been used to trace the mechanical feedback from stars. Much  of the mechanical energy  from fast stellar winds and SNe is expected to be thermalized, at least initially, in diffuse hot plasma \citep[e.g.,][]{Tang2005,Tang2009,vandeVoort2016}. Assuming that the emission arises from this plasma in an optically-thin collisional ionization equilibrium state, one may estimate its mass, energy, and metallicity  \citep[e.g.,][]{Wang2001,Tyler2004,Doane2004,Wang2010,Li2013}. However, such estimation sensitively depends on the {\sl assumed} temperature distribution of the plasma, as well as its volume filling and foreground X-ray absorption properties. This dependence has not been carefully addressed in existing studies. The spectral shape of the diffuse soft X-ray emission in individual galaxies is poorly characterized in general, typically with different combinations of discrete temperature components. Such combinations are highly degenerate and somewhat arbitrarily chosen. As a result, comparison of results from different analyses is often hardly practical or useful \citep[e.g.][]{Kuntz2010}.  

Nearby face-on spiral galaxies provide us with the birds eye views of the diffuse X-ray emission and its relationship to stellar and ISM distributions in galactic disks. Early \chandra\ studies, typically with only moderate exposures ($< 100$~ks), have shown strong correlation between diffuse X-ray and \ha\  emissions \citep[e.g.,][]{Tyler2004,Doane2004,Wang2010}. The tight association  of the X-ray emission with spiral arms is somewhat unexpected, because the mechanical energy output rate from SNe should be nearly constant over a time duration ($\sim 3.2 \times 10^7$ yrs or the lifetime of $\sim 8 \ \rm{M_\odot}$ stars that will eventually end in core-collapse SNe; \citealt{McCray1987}) that is much longer than the lifetime of the stars responsible for \ha\ emission ($\sim 5 \times 10^6$ yrs). This time duration is a considerable fraction of the orbit period of stars around galactic centers. Furthermore, the bulk of the mechanical feedback energy is missing from the X-ray detection \citep[e.g.,][]{Tyler2004,Wang2010,Wang2016}. 
         
 In recent years, deep \chandra\ observations have been obtained for a few nearly face-on galaxies. However, studies based on these data mostly focus on individual discrete sources or their populations. One clear exception is the study of the diffuse X-ray emission in M101 \citep[][KS10 hereafter]{Kuntz2010}. This comprehensive work shows that the bulk ($ \gtrsim 80\%$) of the emission is associated with star forming  regions with ages $\lesssim  20$ Myr, as traced by FUV emission, although the one-to-one correlation between X-ray and FUV ``knots'' is not strong. Furthermore, the X-ray spectrum in the diffuse emission dominant band ($\lesssim 1$~keV) appears to depend on local environment; statistically, regions with higher X-ray intensities tend to have harder spectra.  One would wonder whether or not diffuse X-ray emission in other galaxies behaves similarly and if its origin and evolution could be better understood. 

Similar studies on diffuse X-ray emission from face-on spiral galaxies have been carried out using \XMM\  observations \citep[e.g.,][]{Owen2009,Wezgowiec2020}. Their relatively large field coverage and sensitivity allow to map out the global distribution and to conduct spectral analysis of this emission. Indeed, the studies show $0.2-2$~keV emission permeating the galactic disks. With the limited spatial resolution of  the \xmm\ observations, however, confusion with discrete sources, including background AGNs and clusters of galaxies, is considerable and the separation of spiral arms and inter-arm regions is difficult. A complementary high spatial resolution examination of the X-ray emission remains essential to determine the intrinsic spatial, thermal and chemical properties of the diffuse hot plasma.
                      
\begin{figure*} 
\centerline{
\includegraphics[width=1\textwidth]{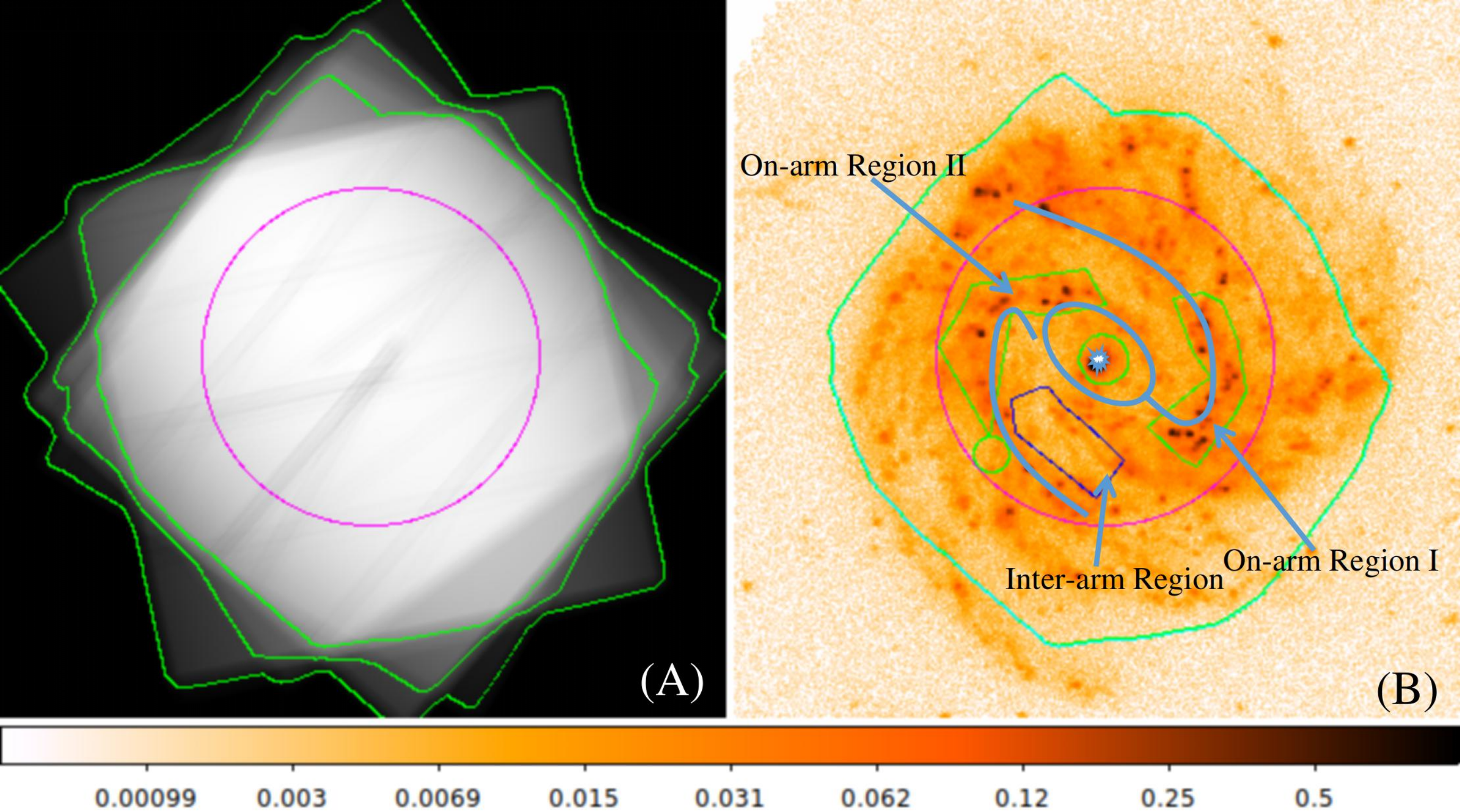}
}\caption{Overview of the \chandra\ overage of \sou: (A) the effective exposure map of the on-axis S3 chip in the 0.45-1~keV band, together with the exposure contours at (0.1, 0.5 and 1) $\times 10^8  {\rm~s~cm^2}$; (B) the $10^8  {\rm~s~cm^2}$ contour (colored in agua), $\sim 50\%$ of the maximum exposure of the map, overlaid on \galex\ FUV image of \sou, together with an ellipse and two curves (all in blue gray) illustrating the central bar and leading edges of the grand-spiral arms of the galaxy. 
The two green circles outline the fields where the very central region of the galaxy and the presence of a background galaxy cluster (see Appendix \ref{a:a1}) dominate the observed X-ray emission; these fields are excluded from our quantitative analysis of the source-removed ``diffuse" X-ray emission. Also marked in both panels is the galaxy's co-rotation radius of $r_c=3.7$~kpc (or 2\farcm8;  \citealt{Foyle2012}; the magenta-colored circle).  The outer disk spectrum of the emission is extracted from the field outside $r_c$, but within the exposure contour, whereas the inner disk spectrum is from within $r_c$; in addition, sample on-arm and inter-arm spectra are extracted from regions outlined by the green and deep blue polygons (see Table~\ref{t:spec-spiral-arms}).
}
\label{f:exp-UV}
\end{figure*}

Here we present a detailed analysis of \sou\ [NGC 5236; centered at R.A. = $13^h37^m00\fs95$; Decl.=$-29^{\circ} 51^{\prime} 55\farcs5$ (J2000.0), NED], based on archival \chandra\ observations (Fig.~\ref{f:exp-UV}). 
Classified as SBc(s) II, at the distance  $D= 4.5$~Mpc \citep[$1^\prime \equiv 1.3$~kpc,][]{Thim2003}
and having a low foreground HI column density (N$_{\rm H}=
4.1 \times 10^{20}{\rm~cm^{-2}}$  \citep{HI4PICollaboration2016},
this face-on grand-spiral galaxy (disk inclination $\sim 15^{\circ}$) 
is well-suited for the study of diffuse X-ray emission and its relationship to the stellar feedback.
Fig.~\ref{f:exp-UV}A  shows the coverage of the \chandra\ observations, focusing on the on-axis back-illuminated S3 chip.
A total of 729 ks exposure is used in the study of discrete X-ray sources of the galaxy, as well as its bright starburst nucleus \citep{Long2014}, which is not included in the present study.  
Following  \citet{Foyle2012}, we assume a trailing spiral structure of the galaxy, rotating in the clockwise direction and having the co-rotation radius of $r_c =3.7$~kpc ({\rm or}~ 2\farcm8). For ease of reference, we trace the leading edges of the two major spiral arms in Fig.~\ref{f:exp-UV}B. The edges,  apparent in high-resolution CO maps, represent the locations of coherent dust lanes seen in optical images. 

\begin{figure}
\centerline{
\includegraphics[width=0.5\textwidth]{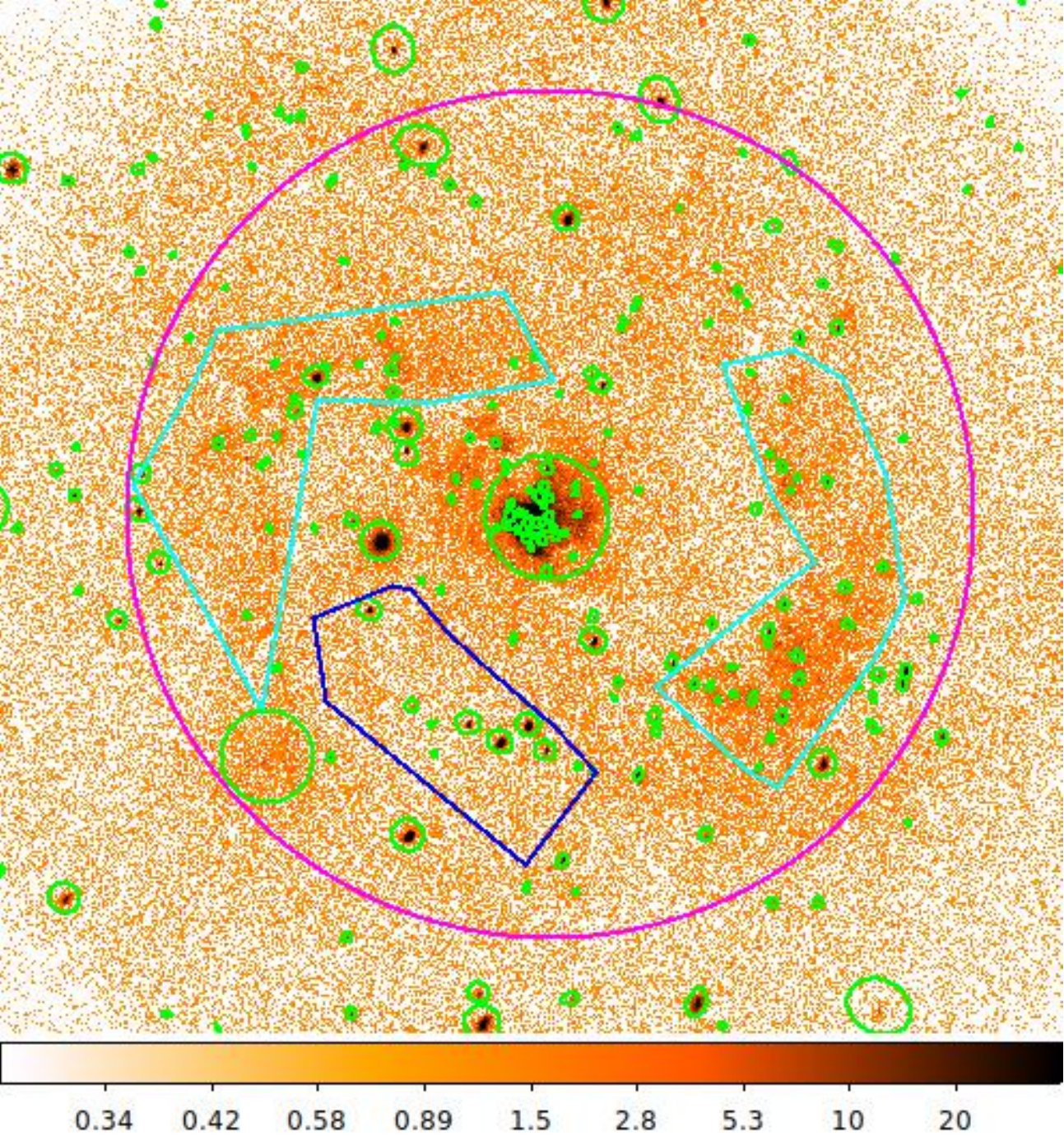}
}\caption{ACIS-S3 0.45-7 keV count mosaic of \sou\ constructed from the \chandra\ observations. Small green ellipses outline the averaged 90\% energy-encircled PSF regions of individual detected sources. The rest of the overlaid regions are the same as those in Fig.~\ref{f:exp-UV}. 
 }
\label{f:x-count}
\end{figure}

The rest of this paper is organized as follows. We describe the \chandra\ observations and data reduction/analysis procedures in \S~\ref{s:obs}. After presenting results of the analysis in \S~\ref{s:res}, we discuss their implications in \S~\ref{s:dis}. Finally, in \S~\ref{s:sum}, we summarize the major results and conclusions of the present study.  Following KS10, we express the intensity of the diffuse  X-ray emission as ${\rm~photons~cm^{-2}~s^{-1}~arcsec^{-2}}$, which will be referred to as surface brightness units (abbreviated as sbu). We present measurement errors (or uncertain intervals) of model parameters at the 90\% confidence.
\section{Observations and Data Reduction/Analysis
}\label{s:obs}
\subsection{\chandra\ data}\label{ss:obs-x}

\subsubsection{X-ray data selection and calibration}
We use the same 10 \chandra\ observations as those used in \citet{Long2014}. They were all taken with the Advanced CCD Imaging Spectrometer (ACIS). 
\sou\ was mainly covered by the back-illuminated S3 chip (Fig.~\ref{f:x-count}). The exposure map shown in Fig.~\ref{f:exp-UV}A includes the effective area of the telescope plus the instrument. For simplicity and uniformity, we chiefly use the data collected by this chip for our on-galaxy data analysis, although the ``off-galaxy'' X-ray background is estimated with the data from the adjacent front-illuminated S2 chip of the observation \# 12994, which had the longest exposure (150~ks) among the observations. 

We reprocess the data with the standard pipeline of the \chandra\ Interactive Analysis of Observations (CIAO; version 4.12 CALDB 4.9.2.1), 
including the light curve cleaning with the routine {\small LC\_CLEAN}.  We construct the mosaic count and exposure maps  in individual bands.  

\subsubsection{Source detection and exclusion}\label{sss:obs-x-sou}
Using the merged maps and their position-dependent and weighted point spread function (PSF), we detect discrete sources, using {\small WAVEDETECT} on the scales of 1.0, 1.4, 2.0, 2.8, 4.0, 5.7, and 8.0 pixels (0\farcs492$\times$0\farcs492 each) and in the soft ($0.3-1.5$~ keV), hard  ($1.5-7$~keV) and broad ($0.5-7$~keV) bands. The detected sources  are merged and are visually examined for their integrity (Figs.~\ref{f:x-count} and \ref{f:x-3color}A). The source detection limit is $\lesssim 10^{36} {\rm~erg~s^{-1}}$ in the 0.35-8~keV band (assuming the distance to \sou), depending on the X-ray absorption, local X-ray background and effective exposure, which are all position-dependent (see also \citealt{Long2014}).

To examine the diffuse X-ray emission, we exclude regions significantly contaminated by the detected sources. For relatively faint sources ($\gtrsim 0.01 {\rm~counts~s^{-1}}$), excluding the 90\% energy-encircled ellipses is sufficient. For brighter source, we manually increase the region sizes to remove visible ring-like features due to the PSF wing. In the present study, we also remove both the central region ($24\arcsec$ radius) of \sou\  and the region ($18\arcsec$ radius)  around a background cluster of galaxies \citep[Figs.~\ref{f:x-count} and \ref{f:x-3color}A; ][see also Appendix \ref{a:a1}]{Long2014}. We estimate the contamination from the PSF spill-out of the discrete sources to be  $\lesssim 5\%$ of the net diffuse flux in the $\lesssim 1$~keV band, which is  mostly considered here; the contamination can be significant, however, in higher energy bands and especially in regions close to relatively bright sources. Another contribution that is not related to diffuse hot plasma is faint X-ray binaries below our detection limit. Their luminosity function is known to be rather flat, although its exact shape at the lower luminosity end remains uncertain, because of the statistical (Eddington) bias in the source detection \citep{Wang2004}, as well as such contamination as supernova remnants and background AGN, which are subject to X-ray absorption in \sou. But according to \citet{Long2014}, the index of the cumulative luminosity function of discrete sources in \sou\ appears to be $\lesssim 0.5$, as expected for low-mass X-ray binaries, which tend to dominate at low luminosities ($\lesssim 10^{37} {\rm~erg~s^{-1}}$) in the \chandra\ energy range. We estimate that the X-ray binary contribution below the detection limit should be smaller than the PSF spill-out contamination of the detected discrete sources.

For the visualization of the diffuse (source-removed) emission only, we fill the holes left from the source exclusion. The filling uses counts randomly selected from events in surrounding regions of the individual sources, using the ciao routine {\small ROI} and {\small DMFILTH}  (Fig.~\ref{f:x-3color}B).  The source-removed regions are not used in the quantitative analysis of the emission (e.g., the construction of the radial intensity profile or spectra).

 \subsubsection{Background Estimation}\label{sss:obs-x-backg}
The background consists of two components:  non-X-ray (cosmic-ray-induced) events and cosmic X-rays. We estimate the non-X-ray background, using data taken while the instrument was in the stowed position, out of the focal plane and under a shield\footnote{http://cxc.harvard.edu/contrib/maxim/stowed/}. 
For each observation, we use stowed events of the same chips to
estimate the non-X-ray background, the level of which is adjusted
so that it matches the count rate actually detected in the 10-12 keV band,
where \chandra\ has virtually no effective area for X-rays. This non-X-ray
background contribution is subtracted from both imaging and spectral
analyses.
 
To study the net X-ray emission from \sou, we also need to estimate the local  X-ray background, which is assumed to be uniform across the field. This estimation is based on modeling the spectrum constructed from counts detected in the ``off-galaxy" S2 chip, after excluding detected sources and accounting the stowed background.  We characterize the X-ray background spectrum with the
spectral model $m_{xb}= APEC+TBABS(APEC+POWERLAW)$, where the names of the spectral model components are in the form used in the spectral analysis software XSPEC (which is part of HEASOFT v6.28). The first component, {\small APEC}, represents the contribution from optically-thin thermal  plasma \citep{Smith2001} of $kT=0.1$~keV (fixed) in the Local Bubble, while the second additive term, $APEC+POWERLAW$,  is for the background emission from more distant diffuse hot gas (e.g., in the Galactic gaseous thick disk and halo) and unresolved point-like sources (primarily AGNs; assumed to have a power law spectrum) and are subject to an X-ray absorption ($TBABS$) due to the Galactic foreground HI column density of $N_H= 4.21 \times 10^{20} {\rm~cm^{-2}}$.  
The solar elemental abundances (aspl in XSPEC) are assumed. These assumptions and physical representations about the individual components are not important, as long as the best-fit model as a whole gives a sufficiently accurate characterization of the {\sl overall}  spectral shape of the sky background. With the  model, we estimate the X-ray background contribution in any on-galaxy region, spectrally or in a broad band, accounting for the area difference in the spectral extraction, as well as the instrument response difference between the two chips (back-illuminated vs. front-illuminated).  Specifically, the X-ray background contribution is $8.65 \times 10^{-10}$ sbu  
in the 0.45-1~keV band -- a band that is optimal for imaging the  emission from diffuse hot plasma, accounting for the uncertainties in the instrument sensitivity and background and the point source contamination (KS10). This contribution is subtracted from our on-galaxy 0.45-1~keV image of \sou.

\subsubsection{Spectral Analysis}\label{sss:obs-x-spec}

 With the limited spectral resolution of the ACIS data, our focus in the present work is to provide a relatively simple and uniform spectral characterization of the diffuse X-ray emission observed in various representative regions  in \sou\ (e.g., Fig. ~\ref{f:exp-UV}B). We adaptively group each on-galaxy spectrum to achieve a $S/N > 3$ per bin, where the specific spectral flux $S$ excludes the non-X-ray background contribution. With the local X-ray background contribution modeled in the off-galaxy field (see \S~\ref{sss:obs-x-backg}), the remaining emission likely represents a combination of  diffuse hot plasma and an accumulated contribution from discrete sources below our detection limit, plus a small residual left from the removal of the detected ones. This latter source contribution, usually insignificant in the soft range ($\lesssim 1.5$~keV), can be important at higher energies and can be adequately modeled with a power law with a fixed photon index $\Gamma= 1.7$ \citep[e.g.,][]{Lehmer2017}.  We assume that the diffuse soft  X-ray emission is dominated by optically-thin thermal plasma in collisional ionization equilibrium and test various models, including single and multiple temperature ones. 
 We find that a (variable abundance) lognormal temperature distribution model of the plasma emission measure ({\small VLNTD} or lognormal plasma hereafter) gives a simple and apparently more physical characterization of the diffuse hot plasma (see more discussion on this in \S~\ref{ss:dis-x-spec} and \citealt{Cheng2021}). The temperature distribution of the emission measure in this lognormal plasma is determined by the two key parameters, the EM-weighted mean temperature $\bar{x}={\rm ln(\bar{T})}$ and the temperature dispersion $\sigma_x$ in logarithm form. However, the fits of the lognormal plasma with a uniform foreground X-ray absorption to the inter-arm and inner disk spectra are not acceptable at $\gtrsim 3\sigma$ confidences. Therefore, we further adopt a multiplicative model {\small LNABS}, which is simply a sum of the foreground X-ray absorption weighted by a lognormal distribution of the absorbing gas column density [see Appendix C of \citet{Cheng2021}], which will be referred as the lognormal absorption model hereafter. This model accounts for the intrinsic absorption in \sou, which is in addition to a fixed uniform Galactic foreground absorption. The inclusion of the model significantly improves the fits to all the spectra (Table~\ref{t:spec-spiral-arms}); each of them cannot be rejected statistically at a confidence $\gtrsim 2\sigma$.

\subsubsection{Imaging analysis}\label{sss:obs-x-image}

\begin{figure*}
\centerline{
\includegraphics[width=1\textwidth]{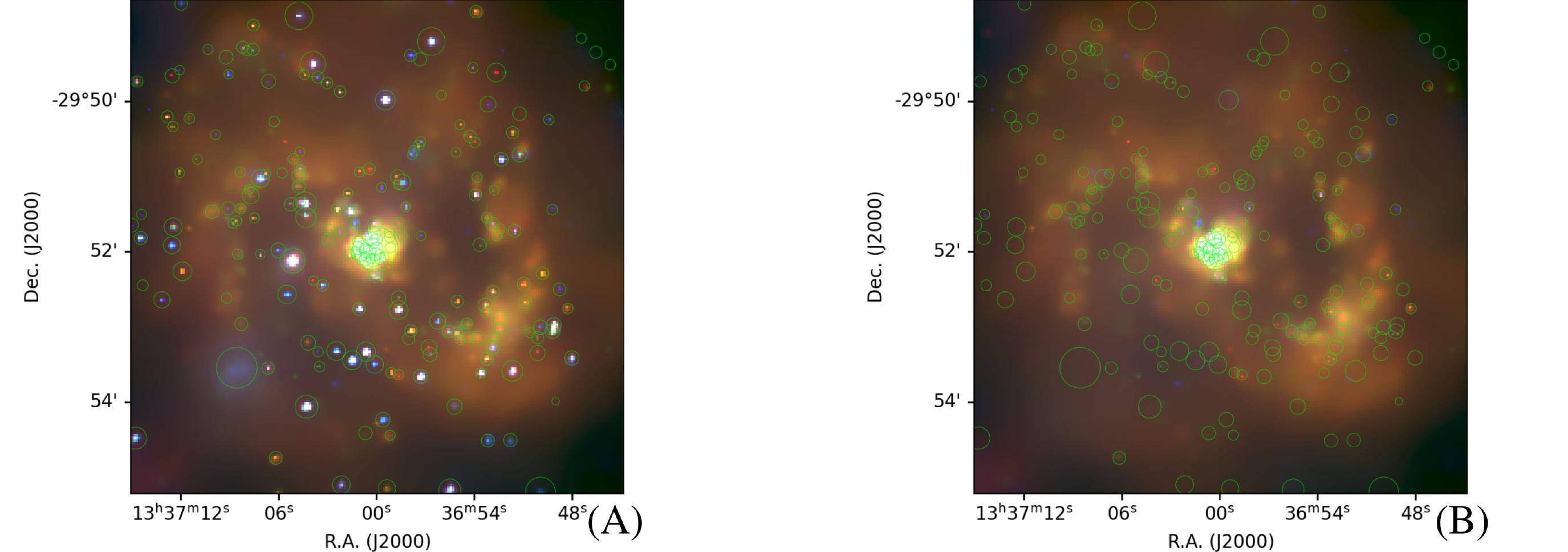}
}
\caption{ (A) 3-color \chandra\  intensity montage of \sou\ in the ACIS-S3 bands: 0.3-0.7 keV (red), 0.7-1.5 keV (green) and 1.5-7 keV (blue). The marked discrete sources are the same as those in Fig.~\ref{f:x-count}.  (B) Same as in (A), but with the sources excised. Images in individual bands are adaptively smoothed. 
}
\label{f:x-3color}
\end{figure*}

\begin{figure*} 
\centerline{
\includegraphics[width=1\textwidth]{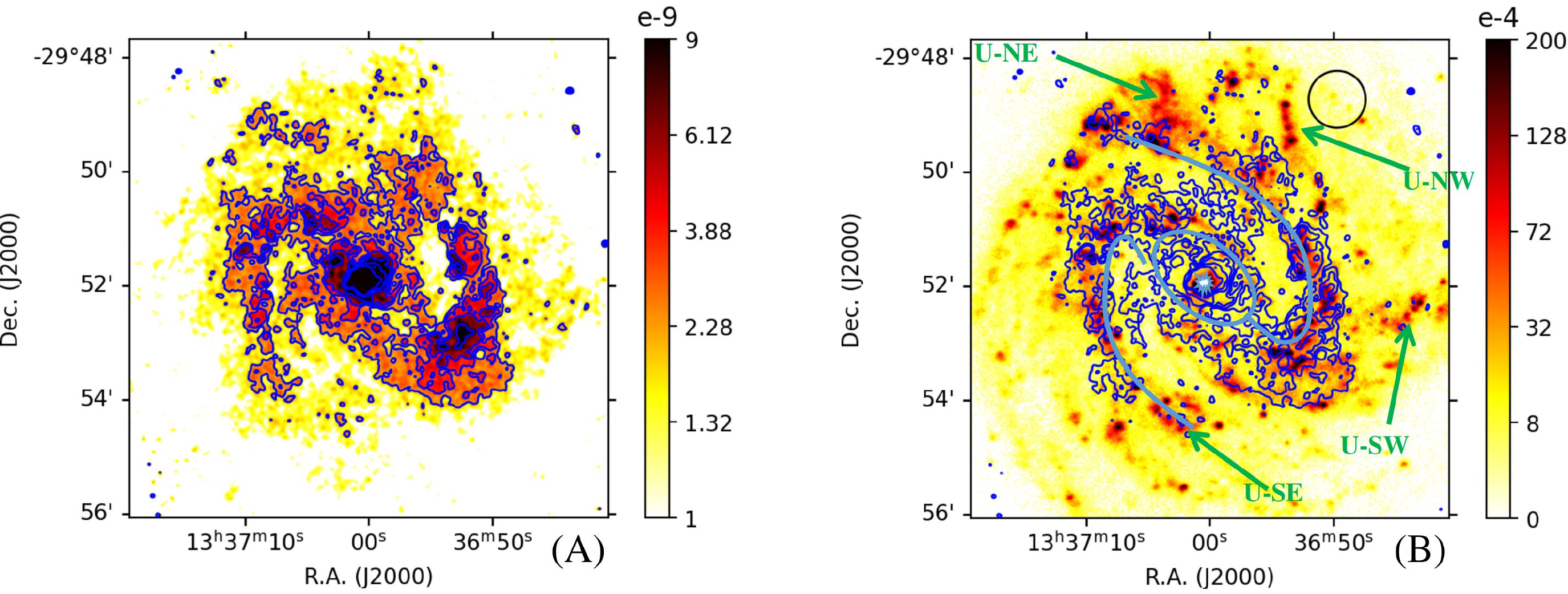}
}
\caption{(A) Intensity image of the diffuse X-ray emission in the 0.45-1 keV band, which is smoothed with a Gaussian of $\sigma =3\farcs9$. The contour levels are at (1.3, 2.6, 4.2, 8.4, 17, and 34) $\times 10^{-9}$ sbu. 
(B) The same contours overlaid on the \galex\ FUV image of \sou. From the image, a local background has been subtracted, which is estimated in the field outlined by the upper right circle. Also marked are a few prominent FUV peaks that do not show corresponding X-ray enhancements in this contour plot. The illustration lines are the same as those in Fig.~\ref{f:exp-UV}B.
}
\label{f:x-dif-gs-UV}
\end{figure*}

We produce the background-subtracted and exposure-corrected intensity images of the  diffuse soft X-ray emission from \sou\ in various bands.  Consistent with the approach used in KS10, our imaging analysis of the diffuse X-ray emission is largely restricted to the ACIS-S3 0.45-1.0~keV range. For presentation only, we smooth the intensity mosaic with either {\small CIAO} {\small CSMOOTH}, which adaptively adjusts the smoothing scale to achieve a signal-to-noise ratio $> 3$ (Fig.~\ref{f:x-3color}), or a Gaussian (e.g., Fig.~\ref{f:x-dif-gs-UV}). 

With the local X-ray background subtracted, we  further construct hardness ratio (HR) maps of the on-galaxy emission, as detailed in Appendix~\ref{a:HR}. The HR is defined as the ratio of the emission intensities in the two bands: I(0.7-1~keV)/I(0.45-0.7~keV), same as that used by KS10. The $0.45-0.7$~keV and $0.7-1.0$~keV bands are dominated by the \ovii\ triplet plus the \oviii\ line  and the Fe-L complex, respectively.  The two images are smoothed with a Gaussian of $\sigma =  
7\farcs87$ before the division to form the HR map. 
However, all our quantitative calculations, including all 1-D presentations (radial and other 1-D intensity profiles), are based on data without any such smoothing.

\begin{figure*} 
\centerline{
\includegraphics[width=1\textwidth]{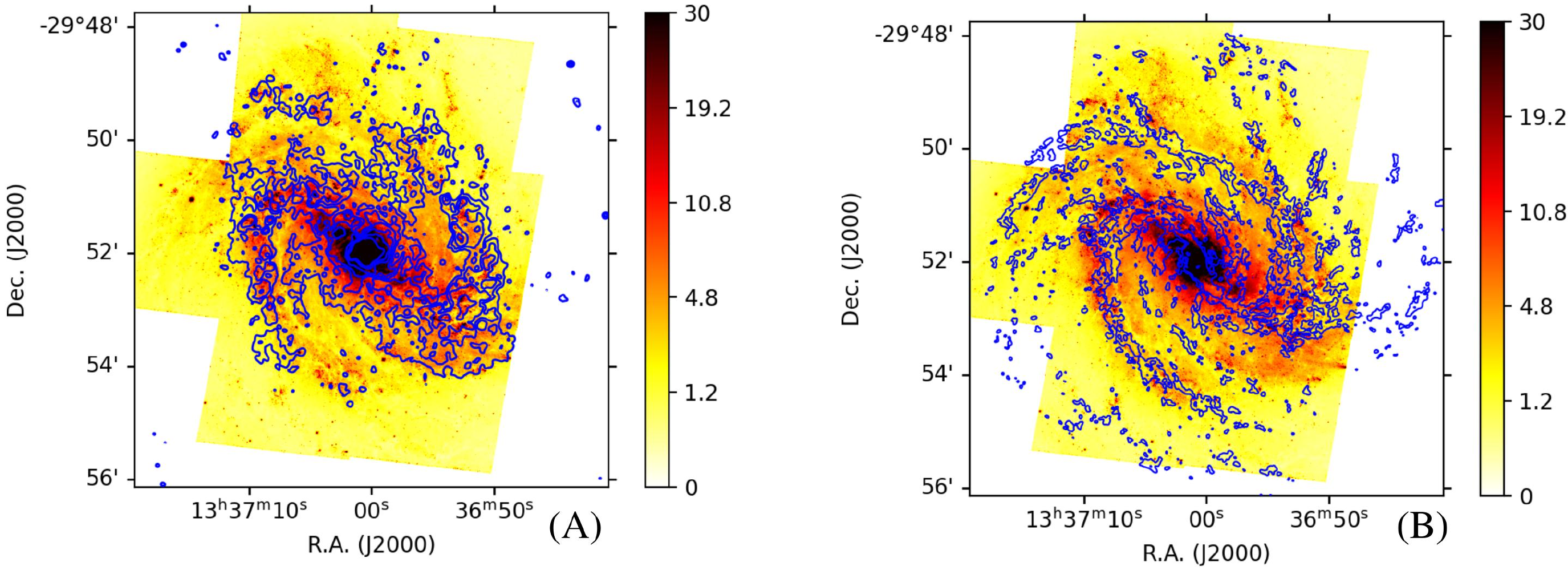}
}\caption{\hst\ \ha\ image overlaid with the X-ray intensity contours same as in Fig.~\ref{f:x-dif-gs-UV}  (A) or with the {\sl ALMA} $^{12}$CO J=2-1 intensity contours at 10 and 50 K~km~s$^{-1}$ (B). These plots demonstrate that the molecular line emission is strongly concentrated along prominent dust lanes or on their slightly downstream sides of the spiral arms. Both the \ha\ and X-ray emissions extend further toward the downstream sides.
}
\label{f:x-Ha-ALMA}
\end{figure*}

\begin{figure*} 
\centerline{
\includegraphics[width=1\textwidth]{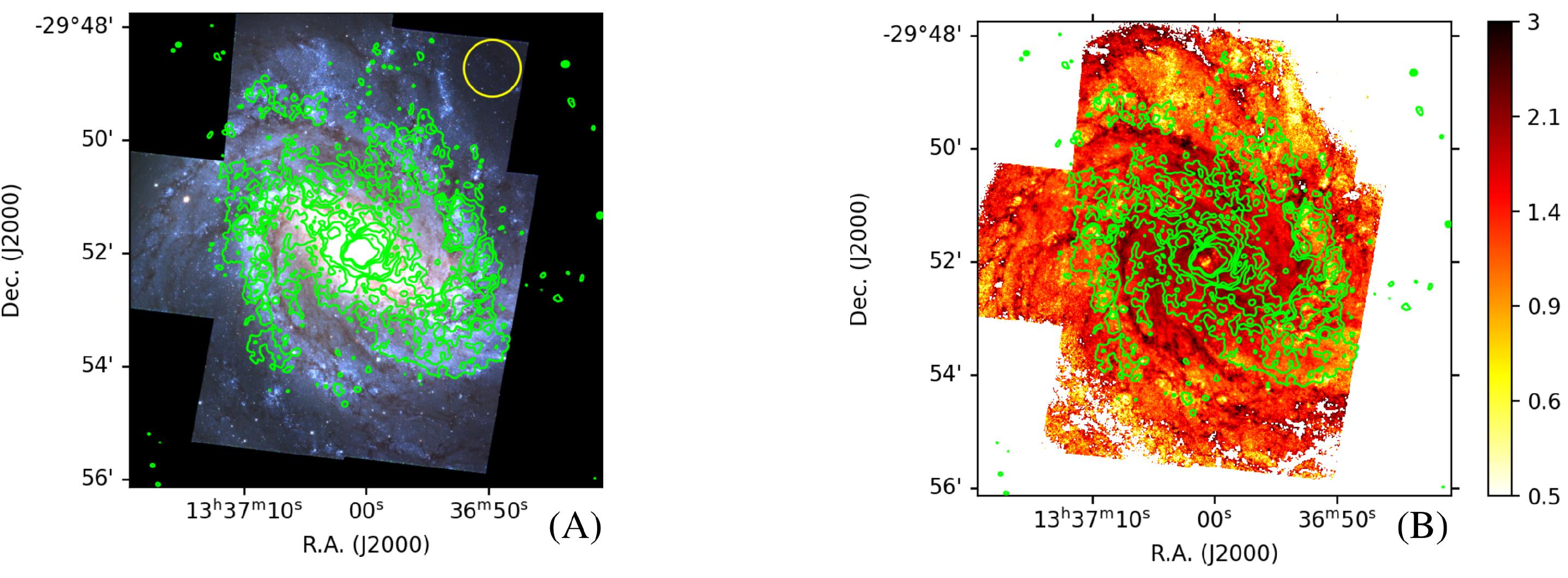}
}
\caption{X-ray intensity contours same as in Fig.~\ref{f:x-dif-gs-UV} are overlaid on the \HST\ 3-color  composite of B-band (blue), V-band (green) and I-band (red) images (A) and the \hst\ B-I color-index image (B). Anti-correlation of the X-ray intensity with the dust extinction is apparent.
The yellow circle in (A) marks the region used to calculate the local background of the \hst\ images.  
}
\label{f:x-3colorHST}
\end{figure*}

\subsection{Auxiliary data}
\label{ss:obs-aux}
We employ multi-wavelength data to trace various stellar and interstellar components of \sou. Some of such data are adopted only for image overlay comparison, while others are included in more quantitative comparisons. Downloaded from the Multi-Mission archive at the Space Telescope Science Institute (MAST)\footnote{\url{https://archive.stsci.edu/missions-and-data/galex}}, the \galex\ FUV image (Figs. ~\ref{f:exp-UV}B and \ref{f:x-dif-gs-UV}B) is mostly sensitive to the emission from young massive stars. The image intensity is in units of ${\rm~counts~s^{-1}~pixel^{-1}}$, where the pixel has a size of 1\farcs5\footnote{\url{http://www.galex.caltech.edu/researcher/data.html}}.
Similarly obtained are also various {\sl HST} WFC3/UVIS mosaick images\footnote{\url{https://archive.stsci.edu/hst/}}, which provide us with  high-resolution optical views of \sou. Specifically, we use the narrow-band F657N (H$\alpha$) image to trace recent massive star forming regions (Fig.~\ref{f:x-Ha-ALMA}) and the broad band F438W (B-band), F555W (V-band) and F814W (I-band) images to mainly probe dust attenuation (Fig.~\ref{f:x-3colorHST}). For example, we construct a color index (CI) map with the F438W and F814W images. To do so, we first estimate the local background (including both sky and instrument contributions) in an ``off-galaxy" region 
 (Fig. \ref{f:x-3colorHST}) and then convert the background-subtracted images to the magnitude ones ($m_{F814W}$ or $m_{F438W}$). The CI in the resultant map ($m_{F438W} -m_{F814W}$) provides a proxy of the dust attenuation and hence the foreground X-ray absorption in the galaxy.

The {\sl ALMA} $^{12}$CO J=2-1 line  emission data on \sou\ are obtained as part of the PHANGS-{\sl ALMA}  survey\footnote{\url{https://www.canfar.net/storage/list/phangs/RELEASES}} \citep{Leroy2021}.  We  here use only the moment0 mosaic map (which combines {\sl ALMA}'s main 12-m array, the 7-m array, and total power observations) to trace dense molecular gas in the galaxy at a  resolution of $\sim 1$\as (Fig.~\ref{f:x-Ha-ALMA}). 

We further use \2mass\ K-band and \wise\ 22\micron\  band images to model the distributions of accumulated stellar mass ($M_*$) and recent star formation rate (SFR) in \sou.  Our interest here is primarily in the relationship of the diffuse soft X-ray intensity to the  feedback from massive stars, which is proportional to the surface (area specific) SFR, while accounting for the contribution from faint sources (mostly cataclysmic variables and other active binaries), which is proportional to the stellar mass. 

\subsubsection{Estimation of the stellar mass X-ray contribution}
\label{sss:obs-aux-sx}
To estimate $M_*$ from the {\sl 2MASS} K-band luminosity $L_K$ 
(both in solar units), we use the relation \citep{Bell2001}
\begin{equation}
    {\rm log}(M_{*}/L_K) =-0.692 + 0.652(B-V),
\end{equation}
which, together with the optical color $B-V=0.59$ (NED)  for \sou, gives the mass-to-light ratio as $0.493 M_\odot/L_{K,\odot}$. We further use the relation \citep{Ge2015} 
\begin{equation}
L_{0.45-1 {\rm~keV}}/M_*= 5.3 \times 10^{27}{\rm~erg s^{-1}M^{-1}_{\odot}}    
\end{equation}
to obtain the conversion to the $0.45-1$~keV X-ray luminosity
\begin{equation}
L_{0.45-1 {\rm~keV}}= 2.6 \times 10^{27}(L_K/L_{K,\odot}) {\rm~erg~s^{-1}}.
\label{e:lx}
\end{equation}
 To apply the above conversion to the 2MASS intensity image ($I_K$ in units of ${\rm DN~s^{-1}~arcsec^{-2}}$), we first calculate the surface magnitude as 
\begin{equation}
m_K=19.9-2.5{\rm log}(I_K),
\end{equation}
 where 19.9 is the magnitude zero point (or the KMAGZP parameter value in the fits header of the image), and then adopt the absolution K-band magnitude of the Sun as 3.28 to obtain the surface luminosity of the image as 
\begin{equation}
L_K/L_{K,\odot}=10^{-0.4[m_k+5-3.28-5{\rm log}(D_{pc})]}.
\end{equation}
With $L_{0.45-1 {\rm~keV}}$ from Equation~\ref{e:lx}, we can express the diffuse X-ray intensity as  $S_{x,*}=a L_{0.45-1 {\rm~keV}}/(4\pi D^2)$, where $a =1.1 \times 10^9 {\rm~sbu/(erg~s^{-1}~cm^{-2}~arcsec^{-2}})$ is the conversion from the energy intensity to the on-axis sbu in the $0.45-1$~keV  band, assuming the best-fit spectral model for the inner disk in Table \ref{t:spec-spiral-arms}. Since we are applying the above derivation to the stellar K-band intensity distribution (i.e., using $I_K$), the units of $L_K$, $M_*$, and $L_{0.45-1 {\rm~keV}}$ should all be considered to be per arcsec$^2$. Putting all these together, we get
\begin{equation}
S_{x,*} =(5.3 \times 10^{-11} {\rm~sbu}) I_K.
\label{e:s-x-star}
\end{equation}

\begin{figure} 
\centerline{
\includegraphics[width=0.45\textwidth]{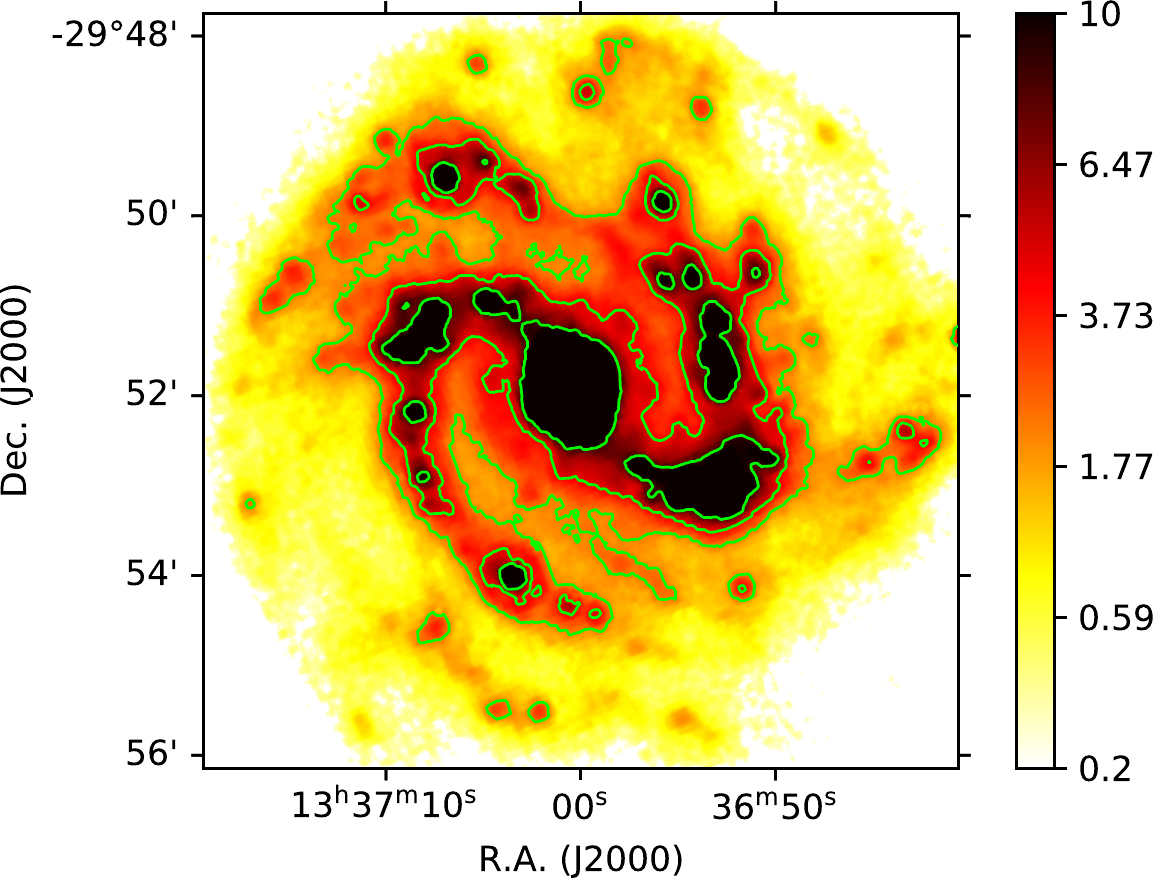}
}
\caption{Processed \wise\ 22\micron\ image  after the subtraction of a local background (25 ${\rm~DN~s^{-1}~arcsec^{-2}}$), as well as foreground stars \citep{Jarrett2019}. The contours are at (0.5, 1 and 2) $\times 4.3  {\rm~DN~s^{-1}~arcsec^{-2}}$ -- the level tracing the surface SFR threshold for outflows (\S~\ref{ss:dis-downstream}).
}
\label{f:wise-image}
\end{figure}

\subsubsection{Surface star formation rate as the stellar feedback tracer}
\label{sss:obs-aux-SFR}
There are various methods to estimate the surface SFR ($\Sigma_{SFR}$) in a nearby star-forming galaxy \citep[e.g., using the FUV+IR emission;][]{Leroy2008,Liu2013}. We here adopt the simply relation between the 22\micron\ spectral luminosity ($L_{22}$) and SFR \citep{Jarrett2013}
\begin{equation}
  SFR({\rm M_\odot~yr^{-1}}) = 7.50 \times 10^{-10}L_{22}(L_\odot), 
  \label{e:SFR}
\end{equation}
where the total solar luminosity $L_\odot = 3.839 \times 10^{33}  {\rm~erg~s^{-1}}$. 
\citet{Jarrett2013} show that the total SFR of \sou\ estimated this way is 2.8 ${\rm M_\odot~yr^{-1}}$, consistent with those from the 24\micron\ (2.8 ${\rm M_\odot~yr^{-1}}$) and the combined FUV+IR (3.2 ${\rm M_\odot~yr^{-1}}$) methods.
Following \citet{Vargas2018}, we here use the above relation to estimate $\Sigma_{SFR}$ (in units of ${\rm M_\odot~yr^{-1}~arcsec^{-2}}$)  of \sou\ from
now the 22\micron\ surface spectral luminosity $L_{22}$. This luminosity distribution can be related to  a \wise\ 22\micron-band intensity image ($I_{22}$ in units of ${\rm DN~s^{-1}~arcsec^{-2}}$), in which, the unresolved stellar contribution in \sou, as well as foreground stars, have been excised \citep{Jarrett2019}. Although the effective angular resolution of this image is quite limited (11\farcs9 FWHM), the availability of the \wise\ all-sky survey makes it attractive to use and easy for comparison with other studies. The resolution is sufficient, considering that the SFR estimation methods are all calibrated on scales $\gtrsim 1$~kpc.

$L_{22} ({\rm L_{\odot}~arcsec^{-2}})$ is related to $\nu I_{22}$ (where $\nu = 1.3\times 10^{13}$ Hz is adopted, corresponding to the central wavelength of 22.8\micron\ of the \wise\ band; \citealt{Brown2014})  via
\begin{equation}
\begin{split}
  L_{22} & =1.3\times 10^{13} \times 5.2\times10^{-5} I_{22} (4\pi D^2)/(10^{23} \times 3.8\times 10^{33})\\
  &=4.4 \times 10^{3}I_{22},
  \end{split}
  \label{e:L22}
\end{equation}
where the term $5.2\times10^{-5}$ converts DN~s$^{-1}$ to Jy (see Explanatory Supplement to the \wise\ 
All-Sky Data Release Products\footnote{\url{https://wise2.ipac.caltech.edu/docs/release/allsky/expsup/}}) and the distance to \sou\ is adopted. Combining the above two equations, we obtain
\begin{equation}
  \Sigma_{SFR}({\rm M_\odot~yr^{-1}~arcsec^{-2}}) = 3.3 \times  10^{-6} I_{22}({\rm DN~s^{-1}~arcsec^{-2}}).
  \label{e:SFR-I22}
\end{equation}
We do not attempt here to match the resolution of the surface stellar mass or SFR model image to that of a particular diffuse soft X-ray intensity image. This latter image is very much limited by counting statistics, especially at a low surface brightness level, and  is typically presented at a spatial resolution lower than those of these relevant multi-wavelength images. In M83, the stellar mass is not particularly structured and contributes weakly to the diffuse soft X-ray intensity. So the resolution should not be a significant issue. Some caution is needed in the comparison of the X-ray data with the surface SFR, which is certainly more structured (Fig.~\ref{f:wise-image}). So when desirable, we also invoke an \spitzer\ 24\micron\ image with a resolution of $\sim 6^{\prime\prime}$ \citep[FWHM; ][]{Dale2009}, downloaded from NED.

\subsection{1-D Decomposition of the diffuse soft X-ray intensity}
\label{ss:obs-rbp}

Using the stellar mass and SFR distributions estimated above, we may  approximately decompose the observed diffuse soft X-ray intensity of \sou\ into two components: 1) the integrated unresolved stellar contribution ($S_{x,*}$); 2) emission traced by $b \Sigma_{SFR}$, where the coefficient $b$ [in units of ${\rm~sbu/(M_\odot~yr^{-1}~arcsec^{-2})}$] is to be determined. This decomposition is conducted with the model radial intensity profile (in sbu):
\begin{equation}
S_{x,m}(r)=S_{x,*}(r)  + b \Sigma_{SFR}(r), 
\label{e:decomp}
\end{equation}
  where $S_{x,*}(r)$ and $\Sigma_{SFR}(r)$ are estimated with the corresponding imaging data (Equations~\ref{e:s-x-star} and ~\ref{e:SFR}). We fit $S_{x,m}(r)$ to the observed radial surface intensity profile of the diffuse soft X-ray emission $S_{x}(r)$ to obtain $b$ .   Again, our focus here is not on an accurate modeling of the diffuse soft X-ray emission, but is instead on its systematic positional deviation from the model to provide insights into the heating/cooling and dynamics of the underlying hot plasma in the disk of \sou.

\section{Results}\label{s:res}

Fig.~\ref{f:x-count} presents the merged count image of \sou\ in the \chandra/ACIS-S3 $0.45-7$~keV band and with the detected sources marked. Fig.~\ref{f:x-3color} shows an adaptively smoothed multi-band composite of the (stowed background-subtracted and exposure-corrected) X-ray intensity images with or without detected sources excised. A Gaussian smoothed version of the intensity image of the diffuse X-ray emission in the $0.45-1$~keV band is given in Fig.~\ref{f:x-dif-gs-UV}. We compare the X-ray intensity contours with those observed in other wavelength bands, including  the \galex\ FUV image (Fig.~\ref{f:x-dif-gs-UV}B), mostly sensitive to massive stars; the \hst\ \ha\ image (Fig.~\ref{f:x-Ha-ALMA}A), tracing recent star-forming regions; and the \hst\ broad band images (Fig.~\ref{f:x-3colorHST}), revealing dust lanes.  Fig.~\ref{f:x-Ha-ALMA}B further compares the \ha\ image with $^{12}$CO J=2-1 intensity contours, depicting a positional relationship of recent massive star-forming regions with dense molecular gas, especially along the spiral arms (see further discussion in \S~\ref{s:dis}). 

In Appendix \ref{a:a1}, we describe the properties of two extended background X-ray sources and their potential utilities for studying \sou. One of these two sources is a distinct background  cluster  of galaxies, which is apparent in Fig.~\ref{f:x-3color}A as an extended hard (bluish) feature projected in the southeastern (SE) part of the galaxy. This cluster is removed in the construction of the diffuse X-ray emission images. The other source is represented by a FR II radio galaxy \citep[e.g., Figs.~\ref{f:f-NW-x-r} and \ref{f:f-spec-maps};  ][]{Cowan1994,Maddox2006}. The X-ray emission from the galaxy forms a linear feature, although its exact extent remains a bit uncertain. In the following, we focus on the diffuse X-ray emission from \sou\ itself.

\subsection{Spatial properties of the diffuse X-ray emission}\label{ss:res-spatial}

\begin{figure*}
    \centering
    \includegraphics[width=1\textwidth]{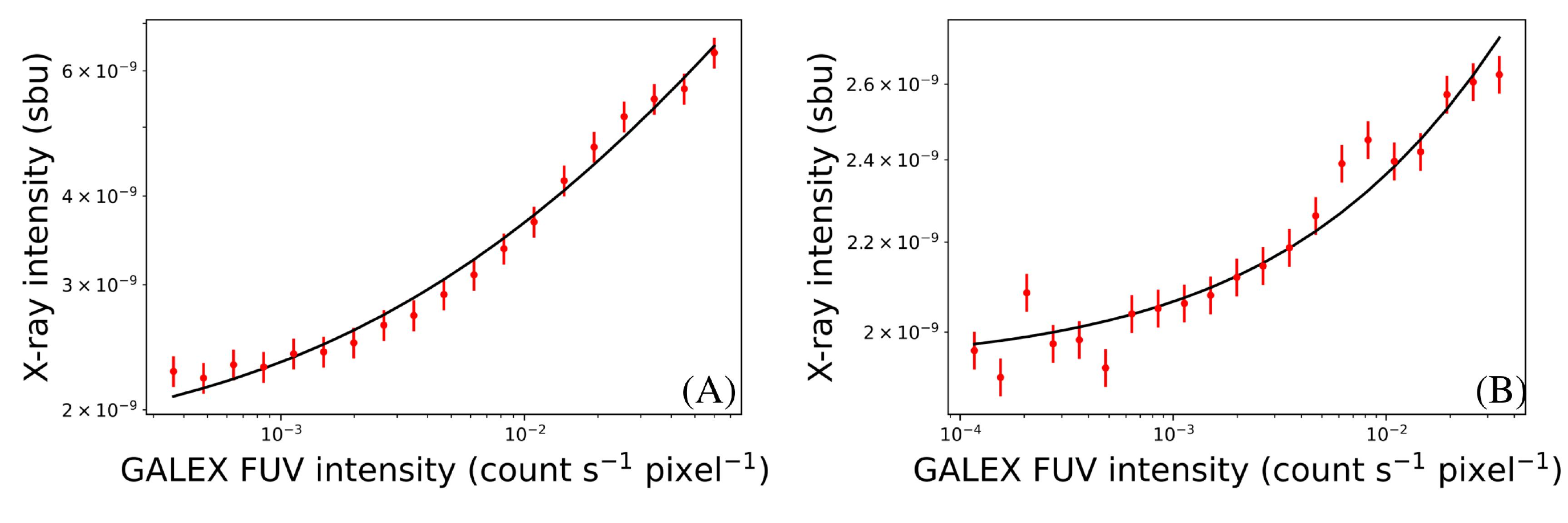}
    \caption{Diffuse soft X-ray intensity     vs. \galex\ FUV intensity: (A) within $r_c$ (but excluding the central region) and (B) between the radius and the $10^8 {\rm~s~cm^2}$ exposure contour (Fig.~\ref{f:exp-UV}). The solid curve represents a fit of the data with $S_{x,m}=(2.0 \times 10^{-8}) I_{FUV}^{1/2}+1.7 \times 10^{-9}$ in (A) and $S_{x,m}=(4.4 \times 10^{-9}) I_{FUV}^{1/2}+1.9 \times 10^{-9}$ in (B).
    }
    \label{f:x-FUV-1D}
\end{figure*}

The grand-design spiral plus bar structure of \sou\ \citep[e.g.][]{Jarrett2013} is apparent in the diffuse X-ray emission image (e.g., Fig.~\ref{f:x-dif-gs-UV}A). Fig.~\ref{f:x-dif-gs-UV}B shows a general correlation between the X-ray and FUV intensities along the spiral arms.  This correlation is more quantitatively demonstrated in Fig.~\ref{f:x-FUV-1D}, where the data are binned according to the ranked values of the x-axis parameter: i.e., in the present case, the diffuse X-ray intensity is calculated with the pixels in each of the \galex\ FUV intensity intervals evenly divided over the presented ranges, beyond which too few pixels exist to conduct a meaningful averaging. The correlation is calculated separately for the two regions: inside or outside $r_c$. Both regions show a consistent square-root dependence of the X-ray intensity on the FUV, as seen in M101 (KS10).

This non-linear dependence also manifests in the spatial correspondence between the diffuse X-ray and FUV intensity enhancements. While they trace each other well at the sharp leading edges of the spiral arms, the X-ray enhancement appears substantially more extended on their downstream sides than the FUV. Furthermore, the X-ray enhancement tends to fill those so-called inter-arm fork/void regions, showing little dusty dense gas (Figs.~\ref{f:x-Ha-ALMA}-\ref{f:x-3colorHST}), and also permeates in the bar (Fig.~\ref{f:x-dif-gs-UV}), where the FUV is not enhanced due to strong extinction \citep{Jarrett2013}. In the outer arm regions (near or outside $r_c$), multiple FUV peaks do not show much diffuse soft X-ray enhancements even at the $1.3 \times 10^{-9}$ sbu level; the most notable peaks (U-NE, U-NW, U-SW and U-SE) are marked in Fig.~\ref{f:x-dif-gs-UV}B.
 Although the background level of the diffuse X-ray emission, which generally decreases with the off-center radius, is a significant factor in determining the weak X-ray appearance of the FUV peaks at the outer radii (see also KS10), other ISM effects also likely play a contributing role (see further discussion in \S~\ref{ss:dis-comp}).

Figs.~\ref{f:x-dif-gs-UV}-\ref{f:x-Ha-ALMA} further demonstrate the relationship among multi-wavelength components in and around the spiral arms.  At the level of $2.6 \times 10^{-9}$ sbu,  bright X-ray knots are generally identifiable with either individual \hii\ regions or larger star-forming complexes chiefly along the spiral arms, as traced by the \ha\ emission. More interesting is the relationship of the diffuse X-ray emission at lower levels (e.g., $1.3 \times 10^{-9}$ sbu) with other components. From the leading edges of the spiral arms toward their downstream sides  (e.g., Fig.~\ref{f:x-3colorHST}), the dust lanes, together with the CO, \ha, FUV, and diffuse X-ray emissions, appear in a sequence, representing the dense dusty gas generation, star formation  and radiation feedback, and stellar mechanical feedback \citep[see also ][]{Jarrett2013,Poetrodjojo2019,Koda2020}. Although these components do overlap with each other, their extents increase in the sequence toward the downstreams. Such differential spatial properties, as well as the sub-linear intensity dependence of the X-ray emission on the FUV, reflect on the evolution of the stellar feedback and its coupling with the ISM across the spiral arms into their downstreams. 

\begin{figure}
\centerline{
\includegraphics[width=0.45\textwidth]{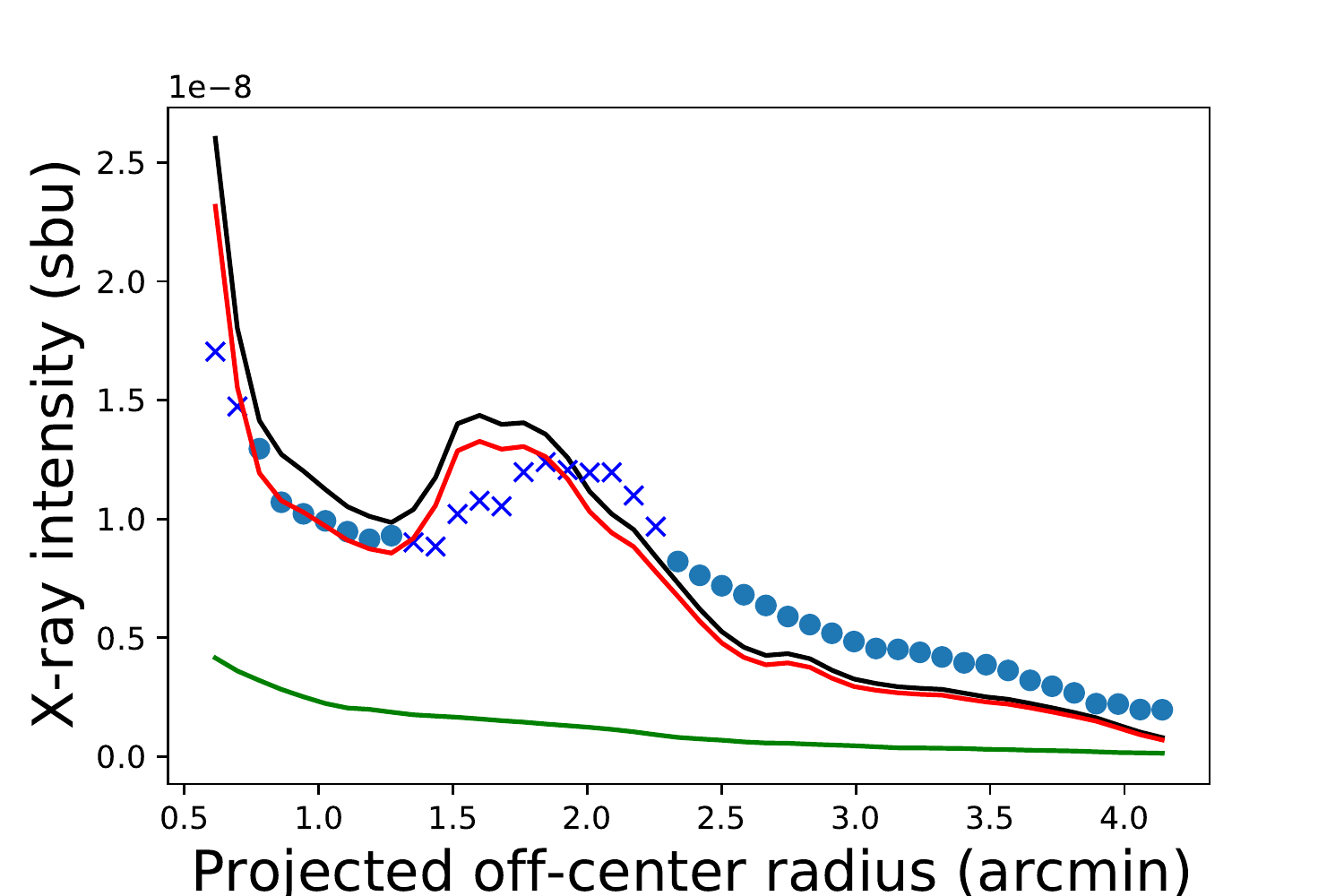}
}
\caption{Approximate decomposition of the radial intensity profile of the diffuse 0.45-1~keV emission (blue points) 
into the contributions due to the stellar mass (green line) and recent SFR (red line). The black line  that fits the data} is the sum of the two contributions. This decomposition is based on a model fit (see Equation~\ref{e:decomp}) to the data points marked as circles; those marked as crosses show too large deviations from the model to be included in the fit. The statistical error of individual data points is only $\sim 2\%$ and is thus not included. 
\label{f:f-rbp}
\end{figure}

\begin{figure*}
\centerline{
\includegraphics[width=1\textwidth]{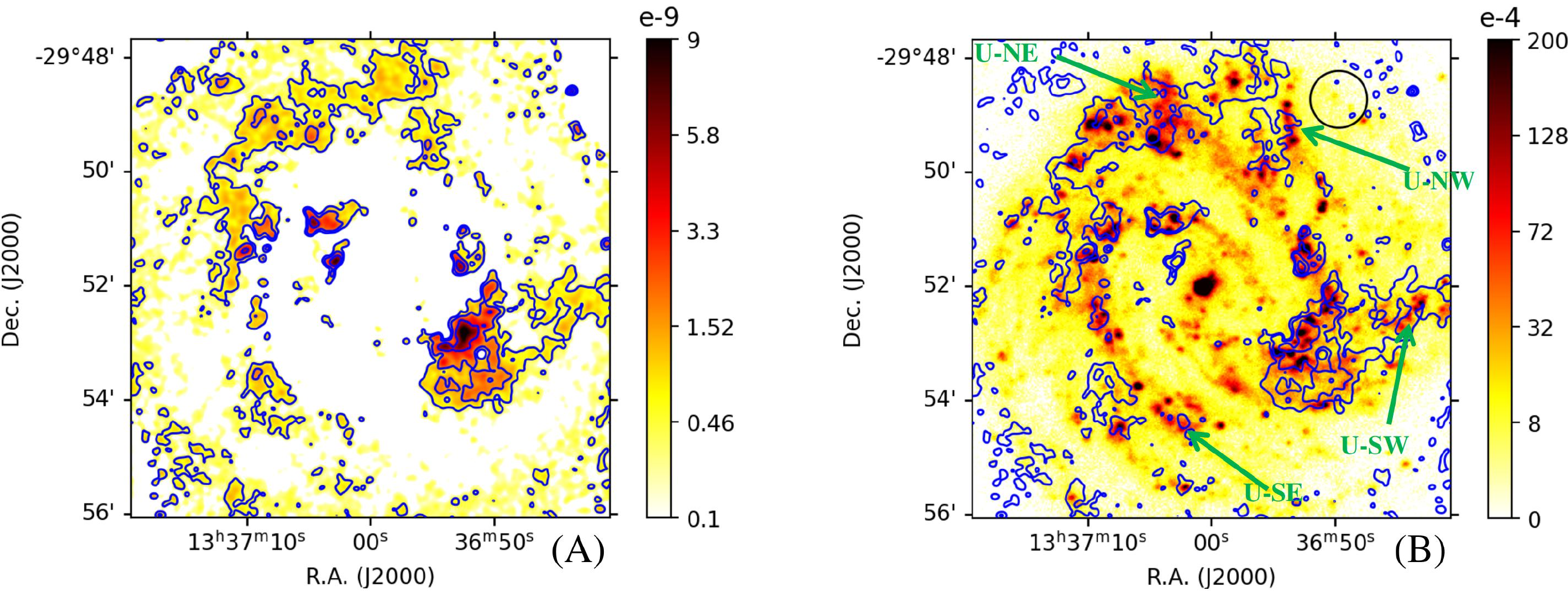}
}
\caption{(A) Residual image of the diffuse soft X-ray intensity after the subtraction of the Model I stellar mass and SFR contributions. (B) \galex\ FUV image as in Fig.~\ref{f:x-dif-gs-UV}. The X-ray intensity contours in both panels are at (5, 15, and  40) $\times 10^{-10}$ sbu.
}
\label{f:res-Model-I}
\end{figure*}

\begin{figure*}
\centerline{
\includegraphics[width=1\textwidth]{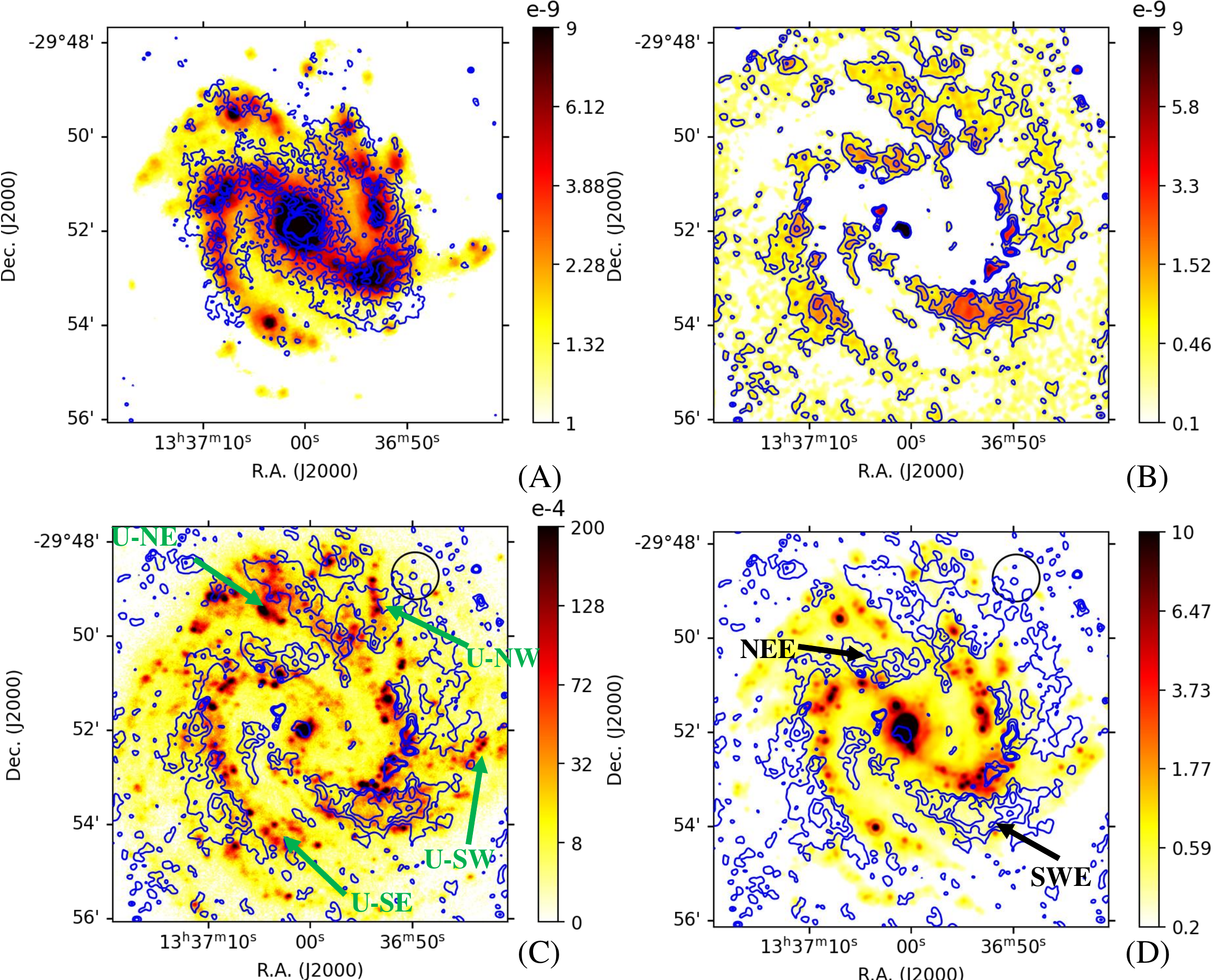}
}
\caption{(A) The same contour as in Fig.~\ref{f:x-dif-gs-UV}, but overlaid on the Model II prediction of the combined stellar mass and SFR contributions. (B) The residual image of the diffuse soft X-ray data  minus the model II prediction in the intensity range of  $(1-90) \times 10^{-10}$ sbu with an overlaid representative residual intensity contours at  $5  \times 10^{-10}$, $1.2  \times 10^{-9}$, and $2  \times 10^{-9}$ sbu.  (C) These same contours overlaid on the FUV image as in Fig.~\ref{f:x-dif-gs-UV}. 
(D)  The same contours overlaid on the 24\micron\ intensity image, together with the (green) 22\micron\ contours at (0.5, 1 and 2) $\times (4.3  {\rm~DN~s^{-1}~arcsec^{-2}})$ for comparison (Fig.~\ref{f:wise-image}). 
Marked are two large-scale X-ray intensity excesses in the downstreams of the spiral arms: the northeast excess (NEE) and the southwest excess (SWE), which can also be appreciated in the azimuthal intensity distribution presented in  Fig.~\ref{f:f-azi}.  }
\label{f:res-2panels}
\end{figure*}

To further examine the differential properties, we probe how the diffuse soft X-ray emission may be globally traced by the stellar mass (dominated by old stars) and the SFR, as described in \S~\ref{ss:obs-aux}. Fig.~\ref{f:f-rbp} shows the radial intensity distribution $S_x(r)$ of the diffuse emission in the $0.45-1$ keV band. While the predicted stellar mass contribution is only $\sim$8\%, the distribution mimics the shape of the SFR contribution, although a relative shift is present, which will be discussed in \S~\ref{ss:dis-downstream}. We here proceed with a fit with Equation~\ref{e:decomp} after excluding data points that show the largest deviations, which gives $b =  8.5 \times 10^{-13} {\rm~sbu/(M_\odot~yr^{-1}~arcsec^{-2})}$.
We consider this fitted (or simply scaled) model as an approximate (or first-order) characterization of the SFR contribution to an accuracy better than $\sim 30\%$ in the disk region concerned here. 

\begin{figure}
\centerline{
\includegraphics[width=0.5\textwidth]{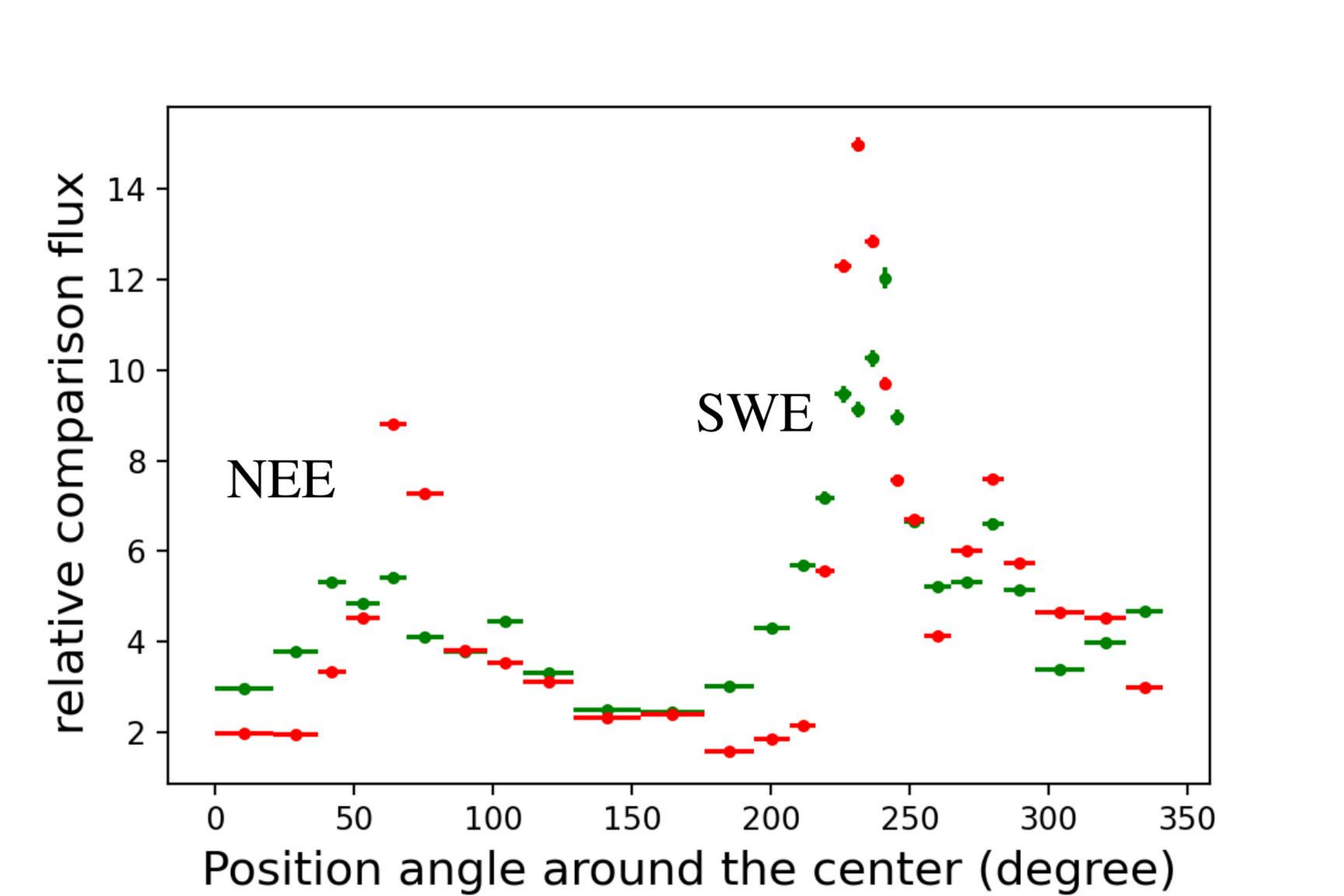}
}\caption{Comparison between the azimuthal intensity distribution of the diffuse 0.45-1 keV intensity (represent by  the green points) and the \spitzer\ 24\micron\ (red points) emissions in the $(0.5-0.8) r_c$ annulus around the center of \sou.  The position angle (x-axis) is in an easterly direction from north. Marked are the two large-scale X-ray  intensity excesses, NEE and SWE (see also Fig.~\ref{f:res-2panels}). 
}
\label{f:f-azi}
\end{figure}

Our main objective here is to facilitate a systematic check of how the observed 2-D diffuse soft X-ray intensity spatially deviates from the SFR distribution. With  the above fitted $b$ value, we may now construct a 2-D model distribution of the emission from an adopted SFR distribution ($S_{SFR,m}$), together with the stellar mass contribution. (\S~\ref{ss:obs-aux}): 
\begin{equation}
S_{x,m}=S_{x,*}  + b S_{SFR,m}. 
\label{e:model-2D}
\end{equation}
We actually build two models. One (Model I) assumes an azimuthally symmetric distribution of the SFR: i.e., $S_{SFR,m}$ in the above equation is represented by the median intensity values in a set of the concentric annuli of the $\Sigma_{SFR}$ image (Equation~\ref{e:SFR-I22}) around the galactic center. This model is used to subtract the contribution (traced by the mean {\sl radial} diffuse 22\micron\ emission) to the observed X-ray emission, in order to reveal the X-ray enhancements that are associated with SFR peaks, including spiral arms, even when they are in outer disk regions ($r \gtrsim r_c$; Fig.~\ref{f:res-Model-I}). The FUV peaks, probably except for U-SE, now appear to show enhanced X-ray emission, in contrast to what is seen in Fig.~\ref{f:x-dif-gs-UV}B. 

To further  demonstrate the morphological shifts of the X-ray enhancements relative to the SFR peaks or spiral arms, we construct Model II by considering $S_{SFR,m}$ in Equation~\ref{e:model-2D} to be the values in the $\Sigma_{SFR}$ image (Equation~\ref{e:SFR-I22}). This model assumes that the 2-D SFR distribution spatially traces the X-ray emission.
Fig.~\ref{f:res-2panels}A shows Model II, compared with the observed diffuse soft X-ray intensity contours   (same as those in Fig.~\ref{f:x-dif-gs-UV}A). By subtracting the model image from the observed one, we obtain the {\sl residual} image of the X-ray intensity (Fig.~\ref{f:res-2panels}B).  Fig.~\ref{f:res-2panels}C-D further compare the residual X-ray intensity enhancement contours with the FUV and \spitzer\ 24\micron\ distributions. The  enhancements (e.g., NEE and SWE) appear systematically away from the leading edges of the major spiral arms toward their downstreams and, in particular, tend to fill local valleys in the 24\micron\ intensity distribution. These trends are quantitatively exaggerated here, because Model II simply assumes a linear scaling of the X-ray intensity with the SFR (in contrast to the square-root dependence as indicated in Fig.~\ref{f:x-FUV-1D}), which leads to  over-subtraction (or under-subtraction) in high (or low) SFR regions. Nevertheless, we believe that this result from a simple and relatively independent  analysis is qualitatively useful.

Fig.~\ref{f:f-azi} directly compares the azimuthal intensity distributions of the diffuse X-ray intensity (which is calculated with the data without smoothing) and the 24\micron\ emission, which similarly traces the SFR at a higher resolution than the 22\micron\ image. The relative X-ray enhancements are apparent at $\sim 0^\circ-40^\circ$ (NEE) and $\sim 180^\circ-220^\circ$ (SWE) on the downstream sides of the NE and SW arms, respectively. 

\begin{figure}
    \centering
    \includegraphics[width=0.5\textwidth]{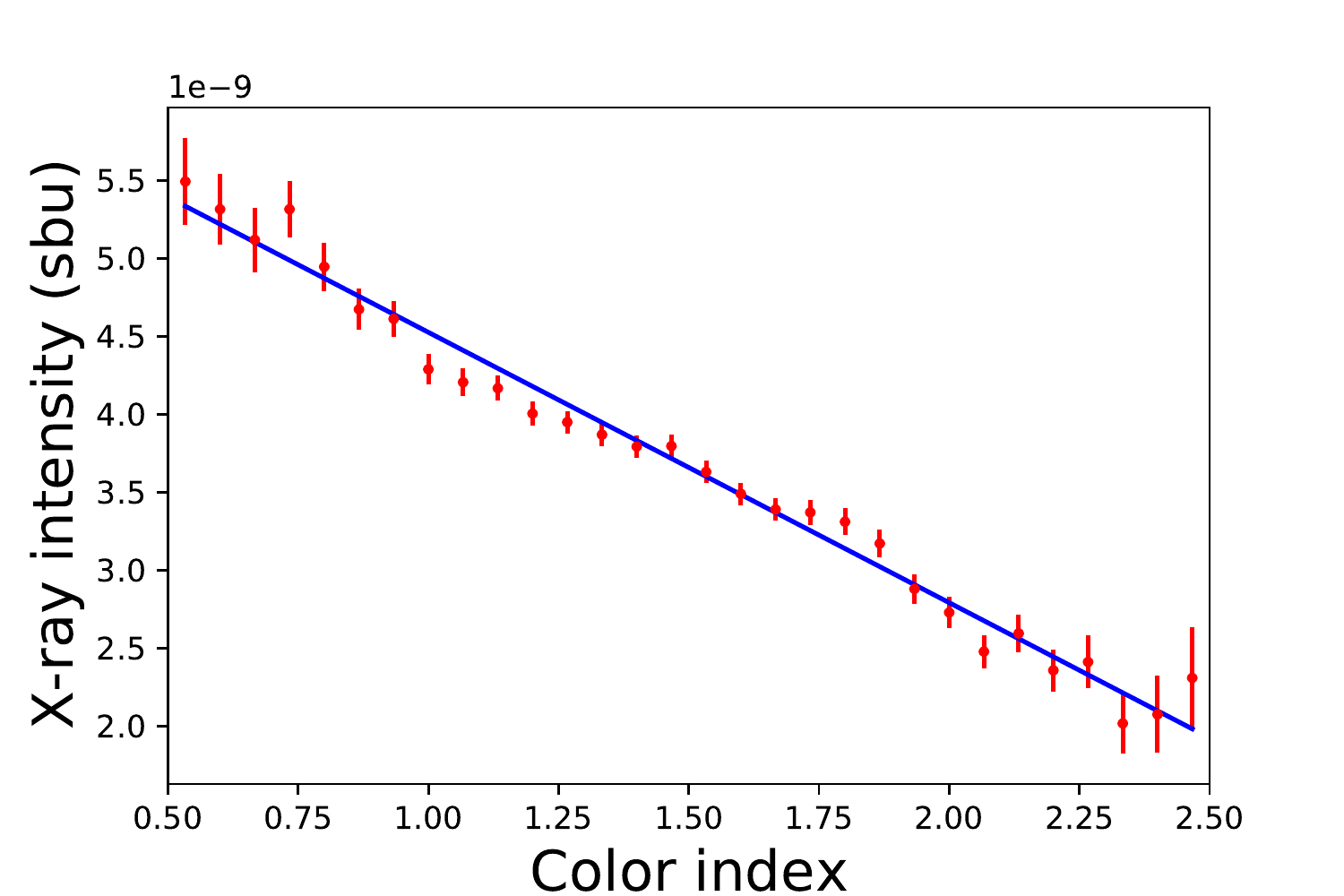}
    \caption{
    Diffuse 0.45-1.0 keV intensity as a function of the CI in the two spiral arm regions (Fig.~\ref{f:exp-UV}B). The solid line is a linear fit with the slope of  $-1.7\times 10^{-9}$ sbu per CI. 
    }
    \label{f:x-CI}
\end{figure}

Furthermore, Fig.~\ref{f:x-3colorHST} shows that the distribution of the diffuse X-ray intensity is affected by the presence of dust lanes. This effect is confirmed statistically by the strong anti-correlation between the X-ray intensity and the CI in the spiral arm regions (Fig.~\ref{f:x-CI}). The anti-correlation is also apparent in inter-arm regions (Fig.~\ref{f:x-3colorHST}). 

\subsection{Spectral Properties of the diffuse X-ray emission}\label{ss:res-spec}

\begin{figure} 
\centerline{
\includegraphics[width=0.45\textwidth]{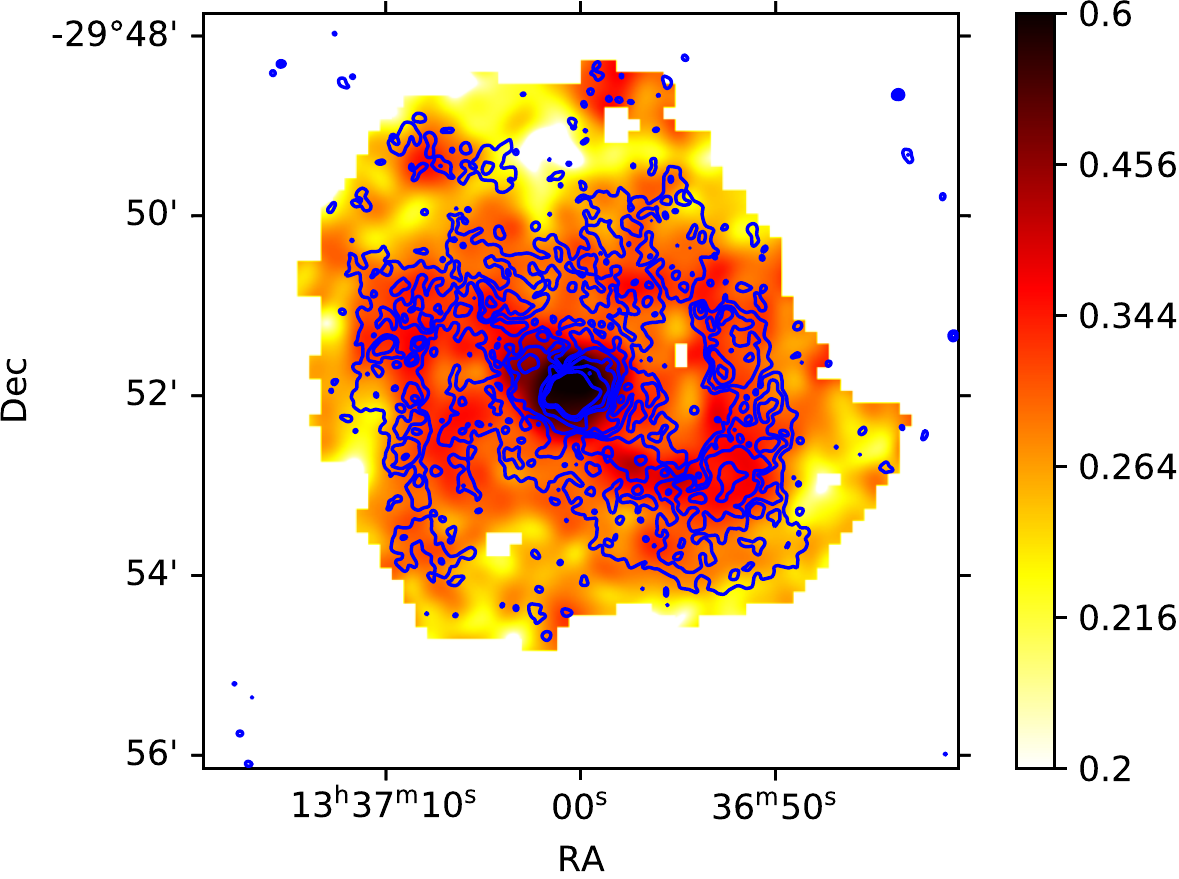}
}\caption{HR [$I$(0.7-1.0 keV)/$I$(0.45-0.7 keV)] map created from the intensity images smoothed with a Gaussian of $\sigma = 7\farcs87$. Only regions where HR $S/N > 3$ are presented here.
The overlaid intensity contours are the same as that in Fig.~\ref{f:x-dif-gs-UV}.
}
\label{f:f-hr-maps}
\end{figure}

\begin{figure*} 
\centerline{
\includegraphics[width=1\textwidth]{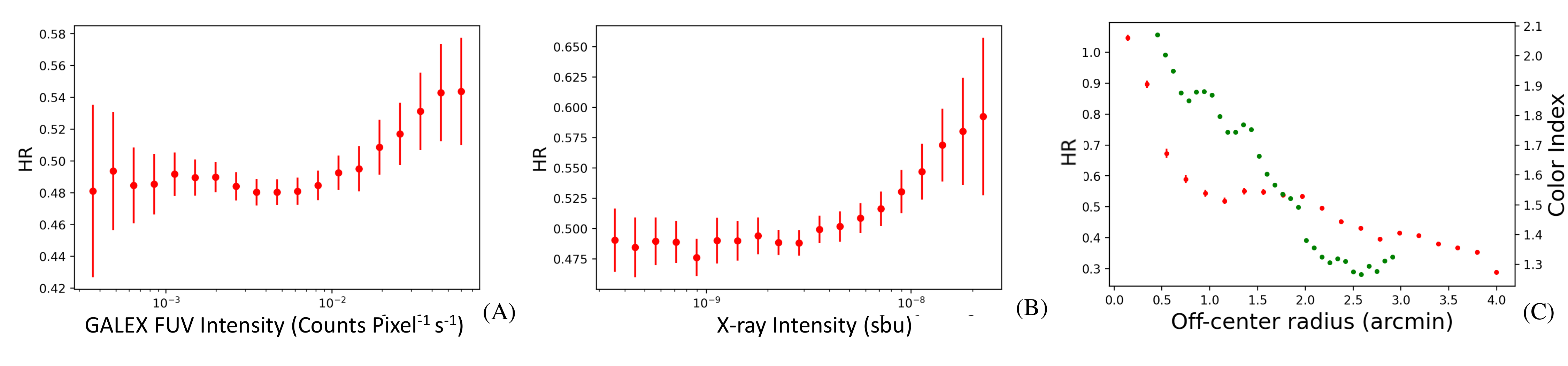}
}\caption{HR as the function of the \galex\ FUV intensity  (A),  the diffuse 0.45-1~keV intensity (B), and the off-center radius (C). For comparison, Panel (C) also includes the radial distribution of the CI (green).}
\label{f:f-hr-prof}
\end{figure*}

Fig.~\ref{f:f-hr-maps} presents the HR map, providing a model-independent overview of the spectral hardness distribution of the diffuse soft X-ray emission across \sou.  The emission is generally harder along the spiral arms than in their downstream and inter-arm regions. This trend is also illustrated in the positive correlation of the HR with the FUV and X-ray intensities (Fig.~\ref{f:f-hr-prof} A-B). Furthermore, the HR tends to decrease with the increasing radius from the galactic center (Fig.~\ref{f:f-hr-prof}C), while locally enhanced along the spiral arms.  These trends suggest that the X-ray emission emerging from active star-forming regions is both enhanced and hardened. 

 The hardness should at least partly be due to relatively strong X-ray absorption, especially in the inner part of the galactic disk. Fig.~\ref{f:f-hr-prof}C includes the radial CI profile constructed from the HST images. Because of their limited field of view, the CI profile extends only to the off-center radius $r \sim 3^\prime$. Nevertheless, the profile shows the flattening of the CI beyond $r \sim 2^\prime$, indicating that the reddening does not change much and probably becomes less important in the outer disk field. In contrast, the HR continues to drop from $r \sim 2^\prime$ to $r \sim 4^\prime$ by $\sim$ 0.3.  This drop is thus most likely due to the temperature decrease of the plasma with the increasing radius. We speculate that the temperature decrease may be largely due to an adiabatic expansion of the plasma from the inner to outer disk, because the X-ray radiative cooling of the hot plasma is inefficient. However, to calculate the expected HR profile for the adiabatic expansion, one needs to know the position-dependent pressure change of the plasma, which is difficult to infer from the existing data with any certainty.

We characterize the spectral properties of the diffuse X-ray emission from selected individual regions (on- or off-arms), as well as from the inner and outer galactic disks (Fig.~\ref{f:exp-UV}B). One may expect that the spectral properties in a small specific region are more uniform than over a large field. Thus our characterization starts with the selected on- and off-arm regions. 

\begin{figure} 
\centerline{
\includegraphics[width=0.45\textwidth]{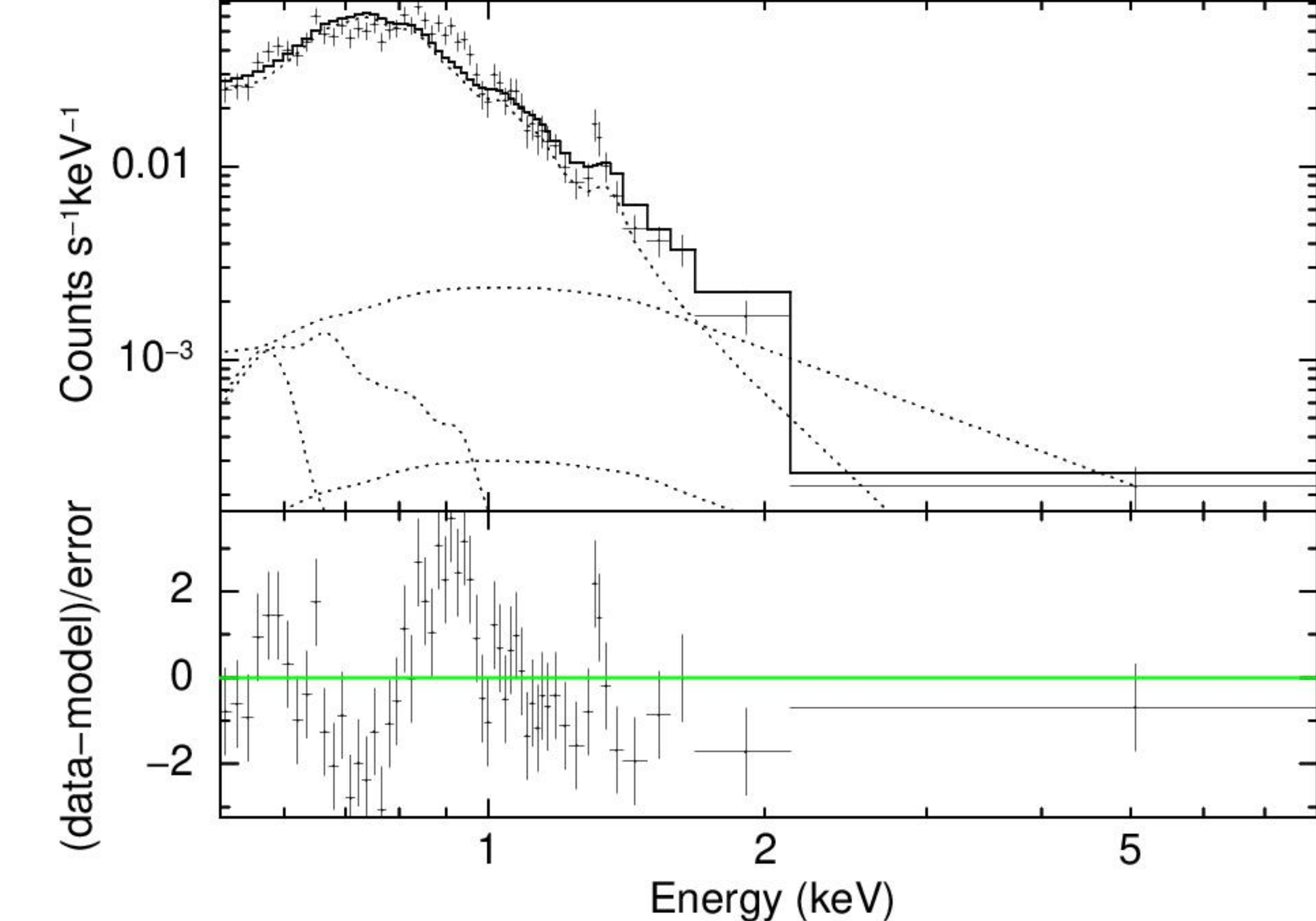}
}\caption{ACIS-S3 spectrum extracted from On-arm Region I (Fig. \ref{f:exp-UV}B), fitted with a single-temperature plasma ({\small APEC}) plus a power law. Individual components (including those associated with the local X-ray background) are separately plotted as dotted lines; the {\small APEC} and power law components are seen as the top two lines at $\sim 1$~keV. This model is an inadequate characterization of the data.}
\label{f:f-spec-spiral-arms-1T}
\end{figure}

We first try fits with a simple {\small APEC} plasma (plus power law) model,together with the foreground absorption column density (N$_H$).
Such fits to the on-arm spectra, for example, give the plasma temperatures of $\sim 0.3$~keV. But the fits with a reduced $\chi^2_r \gtrsim 2$ are far from being acceptable (e.g., Fig.~\ref{f:f-spec-spiral-arms-1T}). Allowing the metallicity to vary does not help much to improve the fitting statistic and leads to a value $\lesssim 0.1$ solar, far lower than what is measured for the ISM in \sou\ and not expected for hot plasma arising from the massive star feedback. 
Fundamentally, the 1-T plasma model spectrum is too peaked, compared to the data.  A plasma of two temperatures (2-T; plus power law) at $\sim 0.2$~keV and $\sim 0.7$~keV gives an acceptable fit with six fitting parameters: two temperatures and two normalizations, plus N$_H$, and the normalization of the power law, while its index is fixed (\S~\ref{s:obs}). 

\begin{figure*} 
\centerline{
\includegraphics[width=1\textwidth]{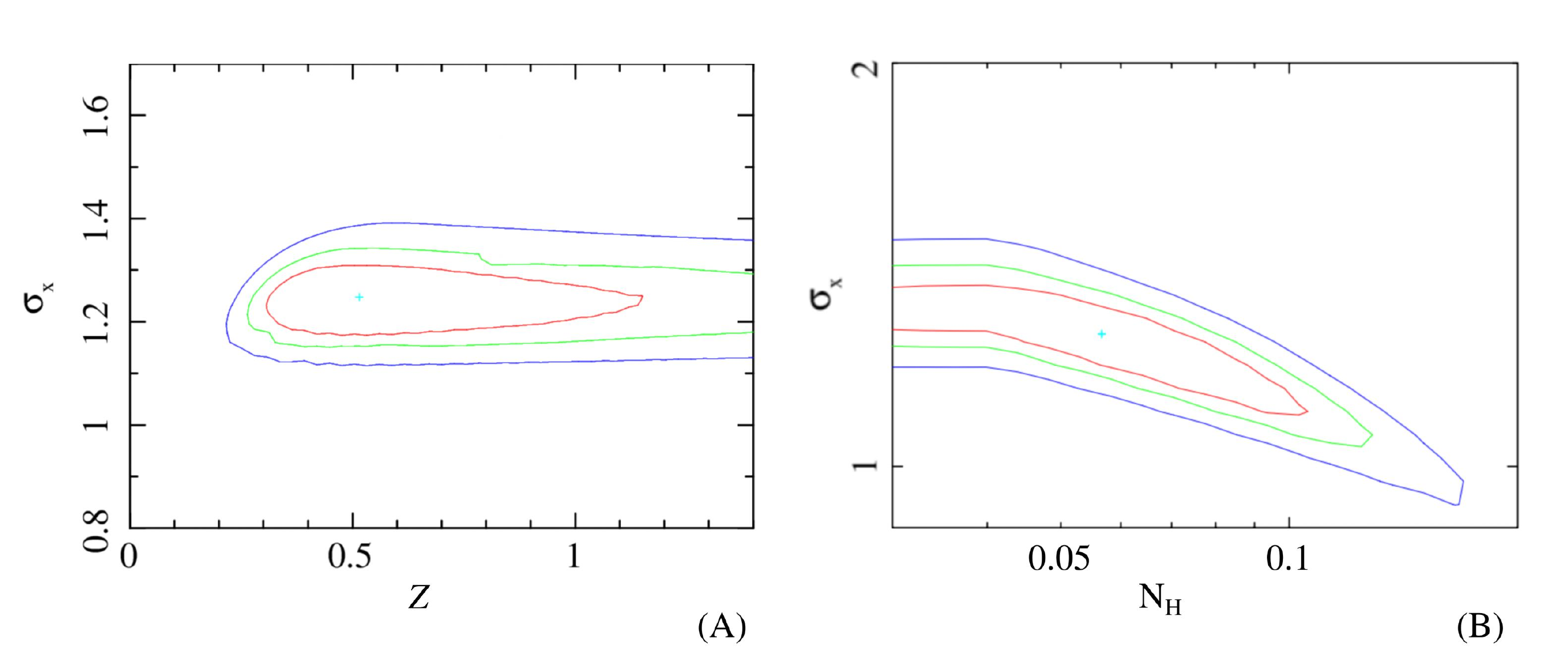}
}\caption{Illustration of the a parameter degeneracy in a spectral fit: the lognormal temperature dispersion ($\sigma_x$) vs. the metallicity ($Z$ in units of the solar value) of the plasma (A) or the foreground X-ray-absorbing  gas column density ($N_H$ in units of $10^{22}{\rm~cm^{-2}}$; B). The contours are at the 68, 90, and 99\% confidences around the best-fit parameters (marked by the plus sign) for On-arm Region I.}
\label{f:f-spec-cont}
\end{figure*}

\begin{figure*} 
\centerline{
\includegraphics[width=1\textwidth]{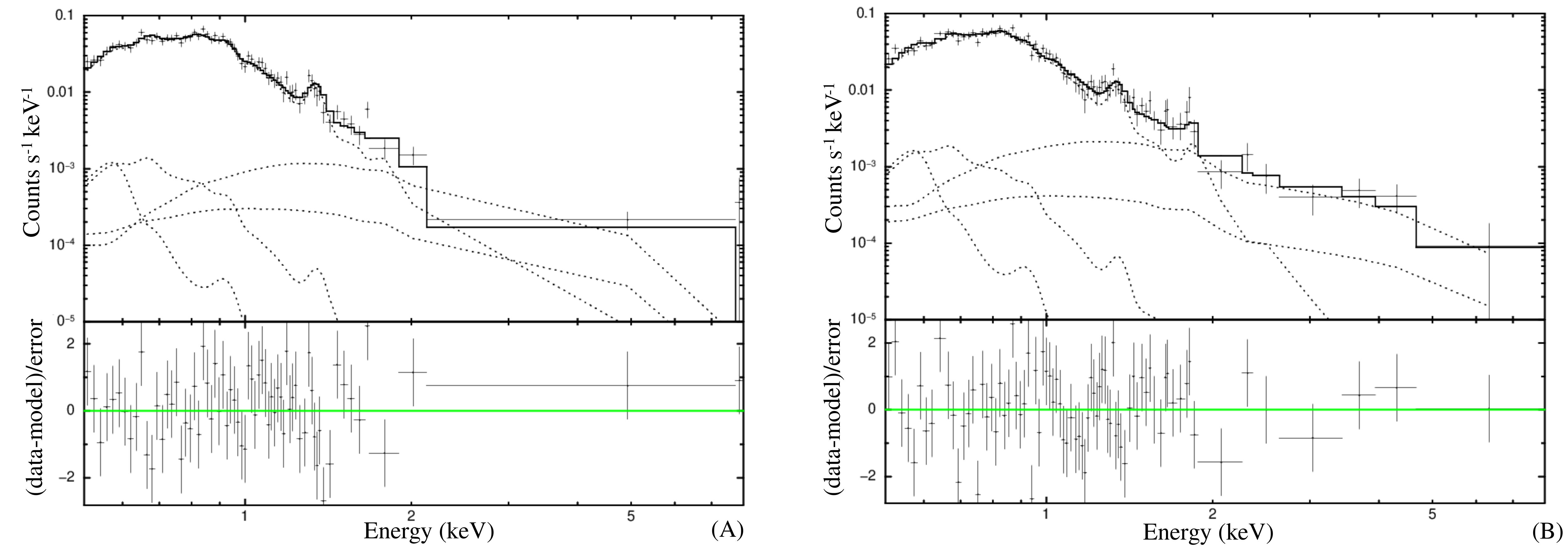}
}\caption{ ACIS-S3 spectra of On-arm Regions I (A) and II (B), fitted with the lognormal plasma plus power law model (Table~\ref{t:spec-spiral-arms}). Individual components (including those associated with the local X-ray background) are separately plotted as dotted lines; the on-arm lognormal plasma and power law components are seen as the top two lines at $\sim 2$~keV. }
\label{f:f-spec-spiral-arms}

\end{figure*}

We find that the lognormal plasma plus power law gives a comparably good fit with fewer fitting parameters (\S~\ref{s:obs}). The quality of the fit and the fitted spectral shape parameters are not particularly sensitive to the metallicity $Z$  as a single parameter (e.g., Fig.~\ref{f:f-spec-cont}A), which is thus fixed to the solar value.  The normalization or emission measure (EM) of the plasma is nearly proportional to $1/Z$ because its emission is dominated by metal lines.
The best-fit mean temperature is anti-correlated with the dispersion $\sigma_x$ and N$_H$ (e.g., Fig.~\ref{f:f-spec-cont}B) and is typically not well constrained in a fit, although a one-sided 95\% confidence upper limit is $\sim 0.15$~keV. To avoid the uncertainty caused by this degeneracy, we simply fix $\bar{T} = 0.1$~keV, consistent with the characterization based on X-ray absorption line measurements of diffuse hot plasma in our Galaxy \citep[e.g.,][]{Gissis2020}. This setup of the model leaves $\sigma_x$ as the only fitting parameter to characterize the spectral shape of the hot plasma component (e.g., Fig.~\ref{f:f-spec-spiral-arms}). As described in \S~\ref{s:obs}, further improvements in all the spectra arise from the use of the lognormal absorption model, plus a Galactic foreground fixed at $4.1 \times 10^{20} {\rm~cm^{-2}}$. Even in this later modeling, which we finally adopt, the total number of fitting parameters ($\sigma_x$ and the normalization of the plasma; $\bar{N}_H$ and $\sigma_{N_H}$ of the absorption; and the normalization of the power law) is one less than the six of the 2-T plasma model mention above.

Table~\ref{t:spec-spiral-arms} summarizes the results from the fits of the final modeling. 
We include the fitted or inferred parameters in Table~\ref{t:spec-spiral-arms}. The modeling for the inner and outer disk spectra with excellent counting statistics, in particular, allows for meaningful constraints on key metal abundances, which also give additional improvements of the fits  (Fig.~\ref{f:f-spec-disk}). The measurements of such individual abundances or their relative values utilize individual spectral line features, which break the above mentioned degeneracy of $Z$ with the normalization of the plasma. The fitted C, O, Ne, Mg, Si and Fe abundances are $4.1^{+1.0}_{-4.1}$ ($3.4^{+1.0}_{-2.3}$),$3.7^{+0.2}_{-2.3}$ ($1.6^{+3.8}_{-1.2}$), $4.9^{+0.5}_{-2.9}$ ($2.0^{+5.0}_{-1.4}$), $3.1^{+0.4}_{-1.9}$ ($1.2^{+2.6}_{-0.8}$), $3.5^{+0.9}_{-1.1}$ ($1.0^{+3.9}_{-0.9}$), $3.1^{+0.3}_{-0.9}$ ($0.9^{+2.3}_{-0.6}$) in the inner (outer) disk, respectively. Other $\alpha$ or Fe-like elements are tied to C or Fe in the fits. It should be noted that
these absolute values of the abundances are systematically uncertain, since they can be sensitive to our specifically assumed spectral models for the individual components (e.g., the index of the power law: the higher the index, the greater the metal abundances). The C abundance, in particular, primarily affects the fitting of the low-energy end of the  spectra and is thus strongly correlated with the foreground absorption.
Considering these statistical and systematical uncertainties, we regard the above fitted abundances as still being consistent with being solar, as has been assumed in the spectral modeling for other regions (Table~\ref{t:spec-spiral-arms}).

Interestingly, the temperature dispersion does not change much from one region to another, all within $\sim 20\%$ of the mean value $\sigma_x \sim 0.65$. Hence no clear evidence is seen here for the cooling of the plasma from the on-arm to inter-arm regions, for example. This is somewhat surprising, as one may expect that the heating of the plasma by CC SNe should be highly concentrated in spiral arms. The measurement of $\sigma_x$ may be complicated by uncertainties in the spectral separation between the lognormal plasma and power law components, as well as the degeneracy with $\bar{N}_H$ and $\sigma_{N_H}$.  These latter two parameters are also strongly anti-correlated with each other (Table~\ref{t:spec-spiral-arms}). It is thus important to consider both parameters in determining the role of the absorption in shaping the observed X-ray spectra.

The spectral fitting results shows that the soft X-ray emission in \sou\ is dominated by the thermal plasma. The power law to plasma flux ratio ($f_{pl}/f_{th}$) in the 0.5-2~keV band is $\lesssim 9\%$.
The flux of the power law component is consistent with the combined contribution from the spill-over of the excised discrete sources and discrete sources that are below our detection limit, including faint HMXBs and young stellar objects as traced by the SFR \citep[e.g.,][]{Bogdan2011,Mineo2012}. 
The power law contribution becomes even smaller in the $0.45-1$ keV band ($\lesssim 5\%$), which we use to map the diffuse soft X-ray intensity distribution.

Using the 0.5-2~keV luminosity of the lognormal plasma of the inner (outer) disk, together with SFR $=1.4~(0.5) {\rm~M_\odot~yr^{-1}}$ in the same region, we estimate $R_{X,SFR} =L_{0.5-2{\rm~keV}}/SFR = 5.2~(4.9) \times 10^{39} {\rm~erg~s^{-1}/(M_\odot~yr^{-1})}$, respectively. 

\begin{figure} 
\centerline{
\includegraphics[width=0.45\textwidth]{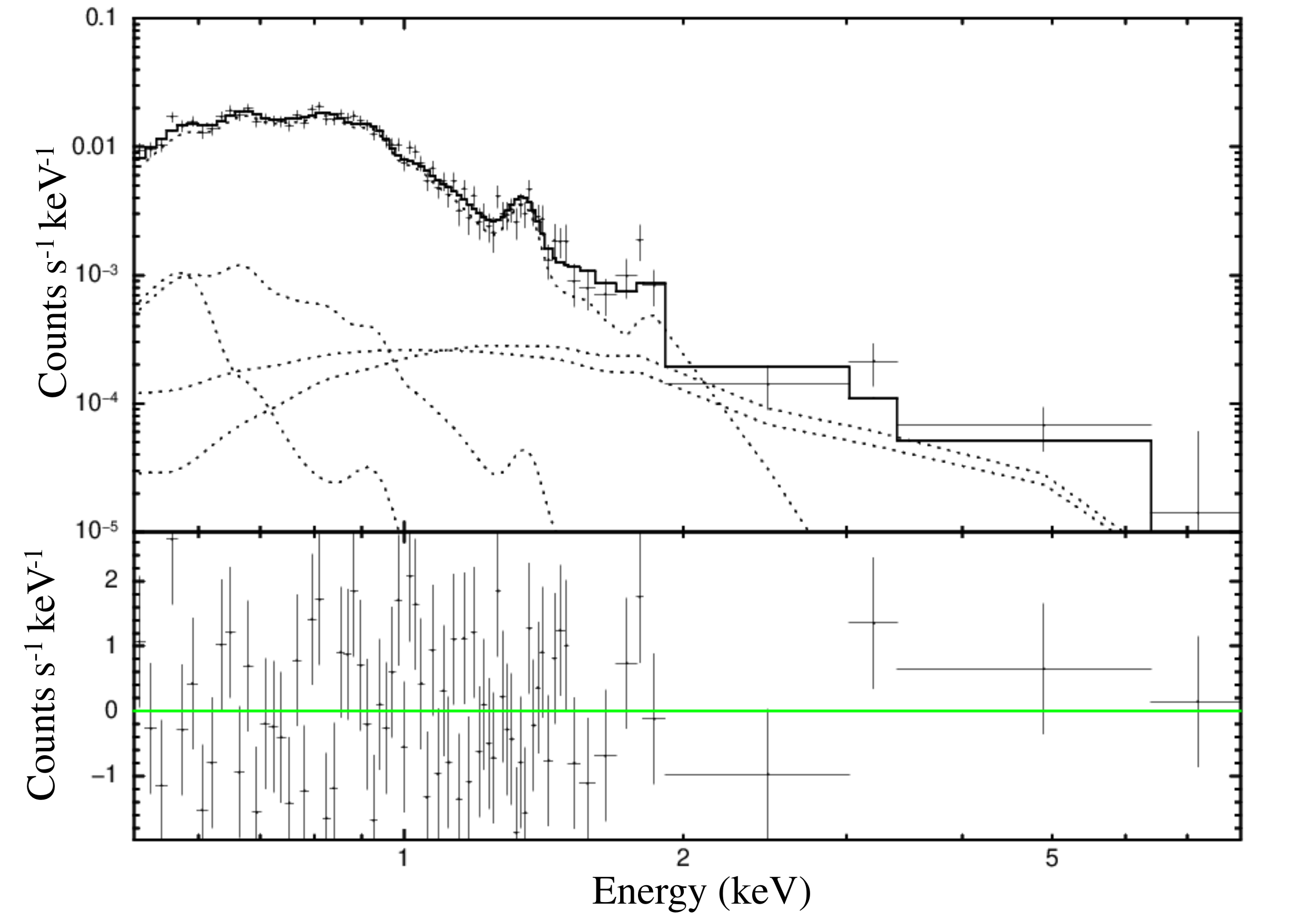}
}\caption{The same as Fig.~\ref{f:f-spec-spiral-arms}, but for a representative off-arm region.
}
\label{f:f-spec-off-arm}
\end{figure}

\begin{figure*} 
\centerline{
\includegraphics[width=1\textwidth]{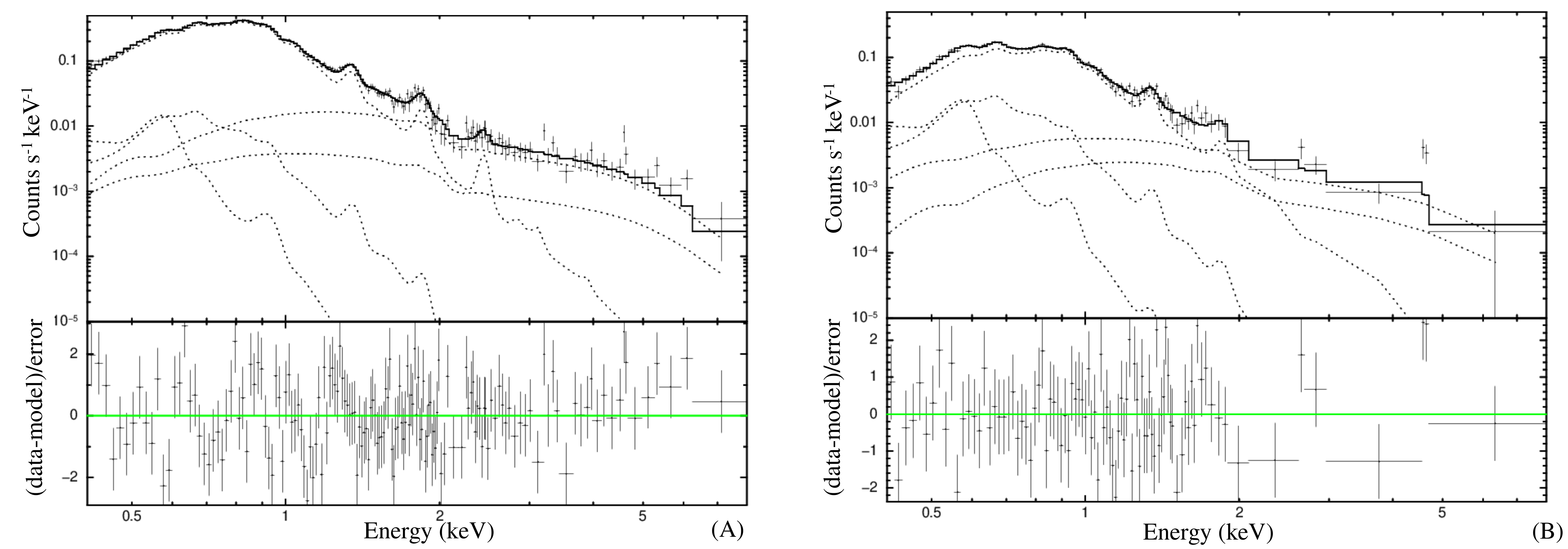}
}\caption{ The same as Fig.~\ref{f:f-spec-spiral-arms}, but for (A) the inner disk spectrum extracted from the region within $r_c$, excluding the central region; (B)  the outer disk spectrum extracted from the region between $r_c$ and the outer exposure contour of $10^8  {\rm~s~cm^2}$ (Fig.~\ref{f:exp-UV}).
}
\label{f:f-spec-disk}
\end{figure*}

\begin{table*}
        {\begin{center} 
        \caption{Spectral fitting results}\label{t:spec-spiral-arms}
        \tabcolsep=0.2cm  
        \begin{tabular}{lcccccccccccccr}
        \\
            \hline\hline
            Regions  &$\bar{N}_H$&$\sigma_{N_H}$& $\sigma_x$  & $K_{th}$ & $K_{pl}$ & $\chi^{2}$/{\rm dof} & Area & $f_{th}$&$f_{pl}$ &$L_{th,x}$ &$P_{th}$ &$E_{th}$&$EM$& $M_{th}$\\
            \hline
            On-arm Region I &26$_{-5}^{+11}$&0.04$_{-0.03}^{+0.20}$& 0.54$_{-0.05}^{+0.05}$ &  6.3$_{-2.4}^{+1.4}$&5.8$_{-2.1}^{+2.1}$&70/63&8195&1.3 &0.05 &5.2 &0.032 & 9.0&1.5&1.6\\
            On-arm Region II &7$_{-1}^{+18}$&0.21$_{-0.06}^{+0.40}$& 0.64$_{-0.07}^{+0.08}$  &  2.8$_{-1.6}^{+0.3}$&8.2$_{-2.0}^{+2.2}$&89/75&8009& 1.3&0.09 &2.6&0.025 &6.8&1.1&1.2\\
            Off-arm Region &6$_{-1}^{+25}$&0.28$_{-0.28}^{+0.20}$& 0.54$_{-0.04}^{+0.04}$&1.7$_{-0.5}^{+0.2}$&1.1$_{-0.6}^{+0.7}$&90/72&4740&0.42&0.01&1.4&0.022&3.6&0.4&0.7\\
            Inner Disk &6$_{-6}^{+12}$&0.18$_{-0.13}^{+0.24}$& 0.75$_{-0.03}^{+0.04}$&7.3$_{-0.6}^{+5.2}$&59$_{-5}^{+5}$&553/508&88669&9.7&0.8&17&0.014&42&1.7&7.4\\
            Outer Disk&10$_{-9}^{+6}$&0.08$_{-0.01}^{+0.01}$&0.79$_{-0.09}^{+0.11}$&3.3$_{-1.4}^{+6.5}$&8.6$_{-5.9}^{+5.7}$&538/508&133820&3.5&0.12&5.2&0.008&37&0.8&3.8\\
            \hline          
        \end{tabular}
       \end{center} }
       \raggedright
        Note: Listed parameters are obtained from the best-fit spectral model consisting of the lognormal plasma and a power law, plus the lognormal absorption, as well as a foreground absorption of a column density fixed to the Galactic value of $N_H = 4.1 \times 10^{20} {\rm~cm^{-2}}$. The mean temperature ($k_B \bar{T}$) is fixed to 0.1 keV. The metal abundances are fixed to the solar values, except for the inner and outer disk regions, for which key elements are fitted (see \S~\ref{ss:res-spec} for the results). $\bar{N}_H$ is  the mean column density of intrinsic absorbing gas  in units of $10^{20} {\rm~cm^{-2}}$, while $\sigma_{N_H}$ is the dispersion of the column density in logarithm  \citep{Cheng2021}. $\sigma_x$ is the dispersion of the temperature in logarithm, while $K_{th}$  and $K_{pl}$ are the normalizations of the lognormal plasma and power law components in units of $10^{-3} {\rm~cm^{-5}}$  and $10^{-6} {\rm~photons~cm^{-2}~s^{-1}~keV^{-1}}$ at 1~keV  \citep{Cheng2021}. $\chi^{2}$/{\rm dof} are the  $\chi^{2}$ value divided by the degree of freedom (dof) of each fit. Also given are the areas (in units of arcsec$^2$) from which the spectra are extracted, as well as derived parameters: the 0.5-2~keV absorbed (no absorption correction) fluxes of the thermal and power law components ($f_{th}$ and $f_{pl}$ in units of $10^{-13} {\rm~erg~cm^{-2}~s^{-1}}$), the  (unabsorbed) 0.1-10~keV luminosity $L_{th,x}$ ($10^{39} {\rm~erg~s^{-1}}$), thermal pressure $P_{th}$ [($f_h h_{\rm kpc})^{-1/2}{\rm~keV~cm^{-3}}$], energy $E_{th}$ [$10^{54} (f_h h_{\rm kpc})^{1/2} {\rm~erg}$], emission measure $EM$ ($10^{63}{\rm~cm^{3}}$) and mass $M_{th}$ (${10^{7}\rm~M_\odot}$) of the plasma.
\end{table*}

With the fitting results, we infer various physical parameters of the plasma, assuming an isobaric state \citep{Cheng2021}.  Our spectral fitting, however, is independent of this assumption, which affects only the interpretation of the fitted parameters and the subsequent inferences. The thermal pressure ($P_{th}$) can then be expressed as 
\begin{equation}
\begin{split}
P_{th} & =\sqrt{\frac{4\pi D^2\eta^2K_{th}}{10^{-14}V_{t}}} (k_B \bar{T}) e^{\sigma_x^2}\\
 & \approx (1.0 \times 10^{33} {\rm~keV~cm^{-3}}) \sqrt{\frac{K_{th}}{V_t}} (k_B \bar{T})_{keV}e^{\sigma_x^2},
\end{split}
\label{e:pressure}
\end{equation}
where $\eta \approx 2.1$ and the distance to \sou\ are adopted, while  $K_{th}$ is the normalization of the lognormal plasma component (Table~\ref{t:spec-spiral-arms}).
The effective total volume ($V_t$) of the hot plasma is $V_t  = f_h V$, where $f_{h}$ is the effective volume filling factor of the plasma in a physical volume ($V$). For a face-on galaxy like \sou, we may approximate $V$  as the area of each region (as projected in the sky) multiplied by the depth $h \approx 1$~kpc along the line of sight: i.e.,
$V \approx (1.4 \times 10^{61} {\rm~cm^3}) A h_{kpc}$, where $A$ (in units of arcsec$^2$, as given in Table~\ref{t:spec-spiral-arms}) is the area from which the spectrum is extracted. 
The total thermal plasma energy is simply $E_{th} =\frac{3}{2} P_{th} V_{t}$, while  
the integrated EM and mass of the plasma can be expressed as
\begin{equation}
EM  = \Big[\dfrac{P_{th}}{\eta k_B \bar{T}}\Big]^2 V_t e^{-2\sigma_x^2},
\label{e:em}
\end{equation}
and
\begin{equation}
M_{th}  = \dfrac{P_{th}\mu m_{p} V_{t}}{k_B \bar{T}} e^{-\sigma_x^2/2},
\label{e:mass}
\end{equation}
where $\mu$ is the mean molecular weight of particles (electrons and ions) and $m_{p}$ is the proton mass.

While the numerical results of these estimations for our spectral analysis regions are included in  Table~\ref{t:spec-spiral-arms}, the above three  equations show how $P_{th}$, $EM$ and $M_{th}$ analytically depend on $\sigma_x$, as well as on those parameters common to the 1-T plasma (e.g., $\bar{T}$ and $K_{th}$). $P_{th}$  in the lognormal plasma is a factor of $e^{\sigma_x^2}$ greater than in an isothermal state (i.e. $\sigma_x= 0$) with a uniform density and at $\bar{T}$. 
This factor corrects for the bias introduced by the EM weighting of the temperature: a large fraction of the volume, occupied by the plasma at the higher temperature (and hence lower density) side of the lognormal distribution, does not contribute much to the observed soft X-ray emission. Similarly, assuming the uniformity and the single $P_{th}$ would over-estimate $EM$ and $M_{th}$ of the lognormal plasma, which results in the correction factors of $e^{-2\sigma_x^2}$ and $e^{-\sigma_x^2/2}$ of their respective expressions.

It should be noted that $\bar{T}$ itself does not necessarily define the observed X-ray spectral shape or even its peak location. In our spectral modeling, much of the X-ray emission from the plasma at the lower temperature side of $\bar{T} \sim 0.1$~keV is absorbed by cool gas along the line-of-sight. The observed X-ray arises mostly from plasma at an intermediate temperature (typically $\sim 0.2-0.7$~keV, as may be characterized by a simple  (1- or 2-T) plasma fit \citep[e.g.,][]{Owen2009,Kuntz2010}. For example,  such a temperature appears at $\sim 1.5\sigma$ above $\bar{T} =0.1$~keV for the lognormal plasma in the inner disk (Table~\ref{t:spec-spiral-arms}): i.e., $T = \bar{T} e^{1.5\sigma_x} =0.32$~keV. The corresponding density at this temperature is a factor of 3.2 lower than at $\bar{T}$.  These dependencies demonstrate the importance of considering the plasma temperature distribution in determining the plasma properties.

We present in Appendix \ref{a:spec_map} the mapping of the spectral parameters (Fig.~\ref{f:f-spec-maps}A-C),  based on the lognormal plasma plus power law model.  Both $\sigma_x$ and plasma normalization maps show a NE-SW elongation and are generally correlated with the spiral arms. The morphological similarity of  these two maps is at least partly due to the high degeneracy between the two parameters in spectral fits. The power law normalization map reveals a linear structure northeast of the galactic center. This NW-SE oriented structure overlaps with the FR II radio galaxy, but seems to extend further to the SE, reaching the eastern spiral arm in projection. It is not clear, however, as to whether or not the entire structure is coherent one or if it is all due to the radio galaxy. There are other apparent peaks in the power law normalization map, which look quite faint, round and extended. They could represent clustered background AGN, which are not detected individually. Although our adopted simple spectral model may be a good approximation globally (due to the averaging), it may not be appropriate on small scales (e.g., due to stochastic spectral realization of a few faint discrete sources, affecting the power law, or local hot plasma or enhanced X-ray absorption, altering the spectral shape in the soft band). With these in the consideration, we find that the fits are generally acceptable across the galaxy (Fig.~\ref{f:f-spec-maps}D), although individual features in the the maps need to be treated with caution. 

\begin{figure} 
\centerline{
\includegraphics[width=0.45\textwidth,angle=-90]{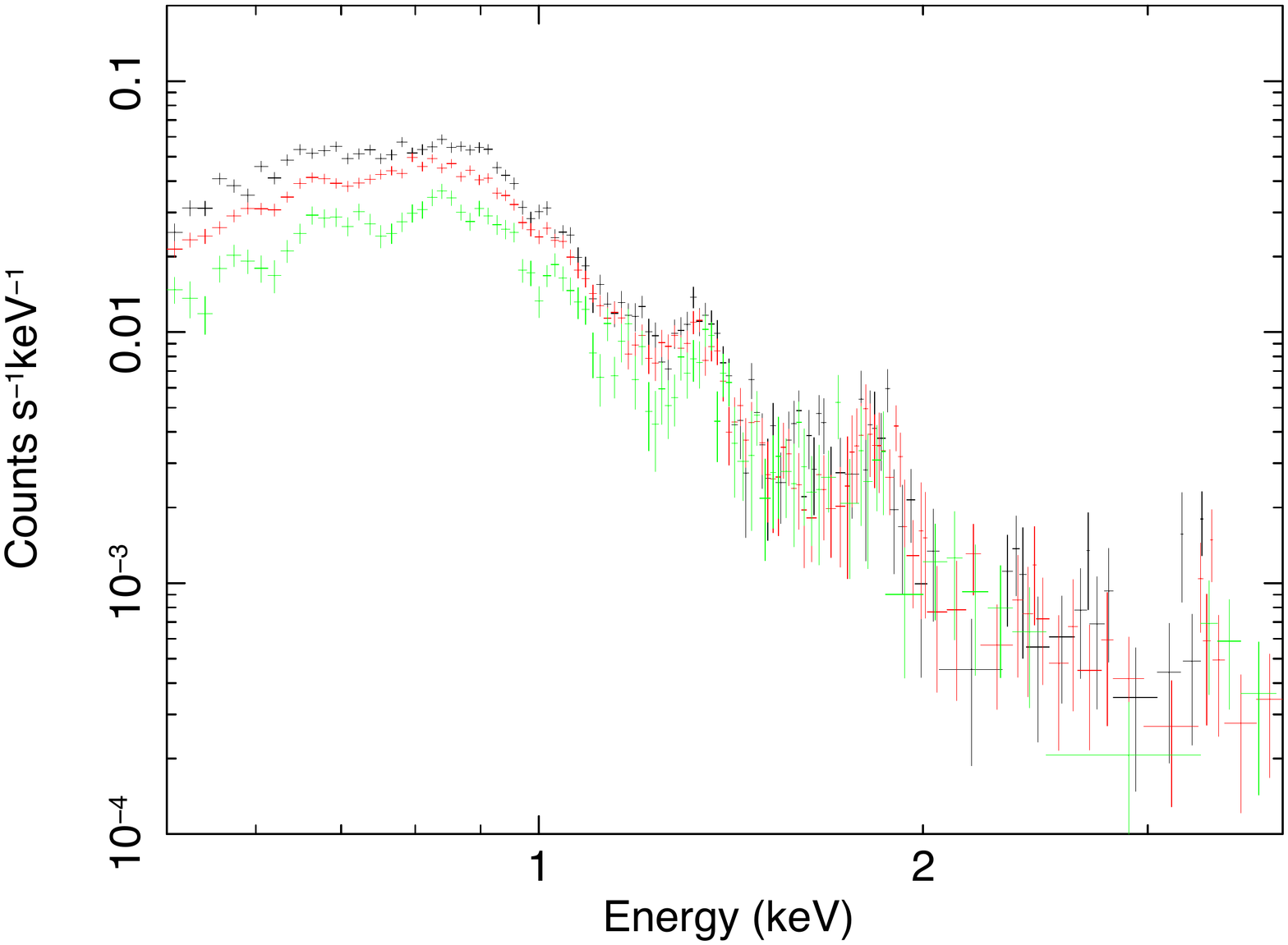}
}
\caption{Comparison of the three spectra extracted in the regions with  ${\rm CI < 1.2}$ (black), between ${\rm 1.2 < CI < 1.75}$ (red), or ${\rm CI} > 1.75$ (green). }
\label{f:f-3spec}
\end{figure}

To check how the spectral properties may be affected by the X-ray absorption traced by the optical CI, we present in Fig.~\ref{f:f-3spec}  three spectra extracted from regions of different CI ranges. Guided by the general trend that the HR increases with the CI, we extract the spectra from the regions with $CI < 1.2$ (CI$_1$), $1.2 < CI < 1.75$ (CI$_2$), and $CI > 1.75 $  (CI$_3$). For ease of comparison, these three spectra have been scaled to the same effective extraction area of 
$10^4$ arcsec$^2$.  The $0.5-1$~keV fluxes in CI$_2$ and CI$_3$ are only 79\% and 52\% of the flux in CI$_1$, whereas the $1-4$~keV fluxes  are consistent with each other. A straight-forward interpretation of these soft X-ray flux deficiencies is that the X-ray absorption is enhanced with the increasing reddening, as traced by the CI. 
We further conduct a joint fit to the three spectra with the lognormal plasma plus power law model, in which the $\bar{T}$, $\sigma_x$ and $\Gamma$ are linked to avoid the large estimation uncertainties of these intrinsic spectral shape parameters which would otherwise be introduced by the degeneracy among them. For the same reason, we use a single $N_H$ to characterize the foreground absorption of each spectrum. However, the quality of this joint fit is poor (judged by its $\chi^2/$dof. So the fitted parameters, especially the absolute $N_H$ values for the individual spectra, may not be quite meaningful, physically. The real situation for the X-ray emission and absorption configuration can be considerably more complicated than what is assumed in this spectral modeling (see more discussion on this in \S~\ref{ss:x-abs}). Nevertheless, the qualitative trend of the X-ray absorption increasing with the CI is quite convincing. 

\section{Discussion}\label{s:dis}

We here synthesize the above results and explore their implications. We start with a comparison of this study on \sou\ with existing ones, especially the one on M101 (KS10), and then discuss major issues related to the production and evolution of diffuse hot plasma, as well as its spatial and thermal structure, and to the X-ray absorption effect of dusty gas in \sou. These discussions provide useful insights into the coupling of the stellar feedback with various phases of the ISM.

\subsection{Comparison with the existing studies on M101
}\label{ss:dis-comp}

This comparison is aimed to enhance our understanding of the results on both M83 and M101. M101 is a prototypical ScI galaxy at the distance of 6.8 Mpc. The work by KS10 is based on similar deep \chandra\ ACIS observations.  Various trends observed in M101 are generally consistent with our results on \sou\ presented above. We find that  the contribution from its old stellar population is small and that the diffuse soft X-ray intensity is strongly correlated with the surface SFR on \sou, consistent with the conclusion reached on  M101 (Figs.~\ref{f:x-dif-gs-UV}B and \ref{f:x-FUV-1D}), showing a similar square-root proportionality. Qualitatively, this sub-linear proportionality is not difficult to understand (e.g., KS10). The FUV emission peaks early in the evolution of a massive star cluster (before the first SN), while the X-ray luminosity of the surrounding hot plasma represents the prolonged heating of the ISM by the mechanical energy feedback, which includes both fast stellar winds and SNe and accumulates with time. The observed diffuse soft X-ray intensity is further affected by the foreground absorption, as well as the evolution of the plasma (e.g., expansion, mass-loading, and potentially cooling). This convolution of the stellar evolution and feedback with the plasma heating and evolution, together with the foreground extinction/absorption, tends to produce a shallow X-ray intensity dependence on the FUV emission. {\sl The square-root dependence, demonstrated here and in KS10, can thus be used as a quantitative constraint on theory and/or simulation for the stellar feedback/ISM coupling.} 

Also consistent between the M101 and \sou\ studies is the apparent weakness of the diffuse X-ray emission from some moderately bright FUV peaks in relatively outer galactic disk regions. As shown in \S~\ref{ss:res-spatial}, part of this weakness is due to the underlying lower diffuse X-ray background emission in such outer regions. Another contributing factor to the X-ray weakness is expected from the generally decreasing metallicity of the ISM with the off-center radius.  According to \citet{Hernandez2019}, the metallicity of \sou\ is $Z \approx +0.20$ dex above the solar value at the galactic center, continuously declines with the radius, and then sharply breaks down to a rather flat distribution at -0.20 dex at $R/R_{25}  \approx 0.4-0.5$, where $R_{25} = 6\farcm44$ is the isophotal radius of the galaxy. Interestingly, the break radius occurs roughly at $r_c$. This decline and drop of the metallicity decreases the plasma emissivity (by  a factor of $\sim 2$). In addition, the density and pressure of the ISM on average decreases with the radius (e.g., Table~\ref{t:spec-spiral-arms}), leading to lower plasma emission measure in superbubbles \citep{MacLow1988}. These factors together should largely explain the relative weak X-ray emission from recently star-forming regions in the outer galactic disk.

M83 and M101 do exhibit interesting differences. While the former contains a distinct bar, which seems to be responsible for the two grand-design spiral arms  of the galaxy. In contrast, M101 shows swirling spiral arms, which are highly asymmetrical and multi-facet and probably result from the galaxy's interaction with neighboring dwarf galaxies. 
This interaction may also account for the asymmetric appearance of M101: The western half of the galaxy appears considerably dimmer than the eastern part in both FUV and X-ray (KS10); Giant H II regions (NGC 5461, NGC 5462, and NGC 5471) are all present in an extended spiral arm on the eastern part of the galaxy. These exceptionally bright regions show high HR in the range of $\sim 0.6-1.1$. There are no such regions in \sou,  the star formation of which is more smoothly distributed  along its two grand spiral arms. The surface SFR is greatly enhanced in the circumnuclear  region of the galaxy, compared to that of M101.  These differences between the two galaxies largely explain their different HR dependence on the off-center radius.  Fig.~\ref{f:f-hr-prof}A exhibits a sharp HR increase (from $\sim 0.55$ to $\gtrsim 1$) from the off-center radius $r\sim 1^\prime$ to the nucleus of \sou. A similar plot for M101 (Fig.~8 in KS10) shows a rather flat distribution of the HR ($\sim 0.4$),  reaching $\sim 0.55$ at the very nuclear region.

In both M101 and \sou, X-ray emission in general appears more extended than the FUV light. This is likely due to the longer cooling time scale of diffuse hot plasma than the lifetime of O and B stars that are largely responsible for the FUV emission.  
However, the confusion among close segments of successive spiral arms along the flow lines prevents a quantitative estimate of the X-ray cooling times for this galaxy. In contrast, the association of the diffuse X-ray emission with the two prominent spiral arms is more distinct in \sou, which makes it an excellent site to study the structure and evolution of hot plasma due to the stellar feedback.

\subsection{Structure of the downstream diffuse X-ray emission in \sou}
\label{ss:dis-downstream}

An important result of our study is the detection of the enhanced diffuse X-ray emission on the downstream side of the spiral arms. There are multiple reasons for the lack of the detection of the downstream enhancement in previous \chandra\ studies \citep[e.g.,][]{Tyler2004}. Almost all of them are based on data that are not sufficiently deep and/or are very much based visual examination of the X-ray emission with various tracers of recent star formation  and/or gas/dust associated with spiral arms. It is also possible that the relationship of diffuse X-ray emission with star forming regions depends on the dynamics and/or trigger mechanism of spiral arms, which may vary from one galaxy to another  \citep[e.g., see the relevant  discussion on M101 and M83 in \S~\ref{ss:dis-comp} and ][]{Shabani2018}. As described in \S~\ref{s:int}, in recent years no much attention has been placed to the global ISM, especially its hot component, on scales of spiral arms, both theoretically and observationally. Detailed studies on such scales will be essential to the understanding of the  stellar feedback and ISM coupling.

The diffuse soft X-ray enhancement observed in the  downstreams of the spiral arms in \sou\ appears to be consistent with a scenario of ISM heating by stellar mechanical feedback. In the downstreams, superbubbles produced by individual massive stellar clusters may have merged to form supergiant bubbles or tunnels of hot plasma on kpc scales comparable to the size of individual spiral arms. Along the way, massive stars are ending their lives and their clusters are being dissolved \citep[e.g.,][]{Lada2003,PortegiesZwart2010}. 
In this scenario, the overall width of the X-ray enhancement associated with each spiral arm is related to the motion of massive stars relative to the spiral arm pattern.  Indeed, the enhancement is largely observed within $r_c$. \sou\ has a flat rotation curve at $180 {\rm~km~s^{-1}}$ and a spiral arm pattern speed of $50 {\rm~km~s^{-1}~kpc^{-1}}$ \citet{Lundgren2004}. So at a representative radius of  $\sim 3$~kpc (where the rotation period is $\sim 1.0 \times 10^8$~yrs), for example, the rotation speed of the stars  relative to a spiral arm may be estimated as $\sim 30 {\rm~km~s^{-1}}$. Assuming a standard initial mass function of massive stars, their mechanical energy input rate (chiefly via  CC-SNe) remains nearly constant over $\sim 3.2 \times 10^7$~yrs -- the lifetime of a 8~$M_\odot$ zero-age main sequence star \citep[e.g.,][]{Leitherer1999}. 
During this time period, such a star will move away from the spiral arm with an offset of $\sim 19^\circ$, which is comparable to the mean value observed for the enhancement. The overall extent of the enhancement is expected to be larger, depending on the time scale of the plasma cooling, expanding, and/or escaping into the galactic halo. 
Therefore, the association of diffuse soft X-ray emission with the downstreams is, at least qualitatively, expected from the heating of the ISM by energetic feedback from massive stars formed in spiral arms. 

The expansion of the hot plasma in and around the galactic disk is consistent with the observed radial distribution trends of the diffuse soft X-ray emission (\S~\ref{s:res}). While the outward shift of the emission relative to the star formation regions (Fig.~\ref{f:f-rbp}) can be a direct consequence of the plasma expansion on the galaxy scale, the decreasing HR with the off-center radius  (Fig.~\ref{f:f-hr-prof}C) is expected from the cooling of the plasma, even just adiabatically. 

Incidentally, the same stellar feedback scenario may explain phenomena observed in other wavelengths. The cosmic-ray generation accompanied by the plasma heating may naturally account for the presence of warm molecular gas, as indicated by the elevated CO J = 2-1/1-0 ratio observed in the same downstreams \citep{Koda2020}.  Magnetic field configuration in the ISM can also be shaped by the feedback. Significant polarized radio emission is detected in the inter-arm regions  \citep{Wezgowiec2020}, which are away from the downstreams where the enhanced diffuse X-ray emission is observed. This is consistent with the hypothesis that the generation of the hot plasma is largely responsible for the turbulence, which tends to wipe out organized magnetic fields in the ISM \citep[e.g.,][]{Bacchini2020}.

\subsection{Validity and implication of the lognormal models of the plasma and absorption}\label{ss:dis-x-spec}

With a continuous temperature distribution, the lognormal plasma characterization should be more physical than those involving two or more discrete temperature components \citep{Cheng2021,Vijayan2021}. The lognormal distribution is expected from the central limit theorem for a variable (e.g., temperature) characterizing the accumulation of many independent multiplicative random processes such as repeated shock-heating of the plasma, as demonstrated in various hydrodynamic simulations (e.g., \citealt{Frank2014}). 
Of course, a continuous temperature distribution is also expected from plasma cooling, radiatively or via turbulent mixing \citep[e.g.][]{Lancaster2021}. In any case,  the lognormal plasma characterization includes or extends the single-temperature model with only one additional parameter (the dispersion of the temperature in logarithm). In comparison, the  modeling with plasma of multiple temperatures would add more parameters, which typically have little actual physical meaning  and are forced by the discrete setup \citep[e.g.,][]{Gayley2014,Vijayan2021}. 

 Similarly, the lognormal characterization of the X-ray absorption ({\small LNABS}) should also be more realistic than the assumption of a single foreground absorbing gas column density, as commonly used. Strong X-ray absorption variation is clearly seen across individual star forming regions \citep[e.g.,][]{Cheng2021}. 
Such variation is expected to be greater across much of a star forming galaxy, including both on- and off-arm regions.

Indeed, as we have demonstrated in \S~\ref{ss:res-spec}, our spectral analysis shows the effectiveness of the lognormal characterizations of the plasma temperature and absorbing gas column density distributions. In particular, no evidence exists for a preferred peak in the temperature range that the X-ray data are sensitive to ($\gtrsim 0.1$~keV). The single temperature plasma fit is not acceptable when the S/N of a spectrum is good (e.g.,  Fig.~\ref{f:f-spec-spiral-arms-1T}).  
The lognormal plasma with the mean temperature fixed to 0.1 keV (thus the same number of fitting parameters as the 1-T model) gives much improved fits to all the spectra that we have analyzed.  The application of the lognormal absorption, instead of a single absorption, further improves the fits (\S~\ref{ss:res-spec}). Our results also show that the combined applications of the lognormal plasma and absorption models lead to the estimates of the metal abundances that are consistent with or slightly higher than the metallicity measured from the integrated UV light of star clusters \citep[\S~\ref{ss:dis-comp}; ][]{Hernandez2019}. In contrast, X-ray spectral fits with plasma of discrete temperatures tend to give estimates of the metal abundances that are substantially lower than measured values for ISM or young stars \citep[e.g.,][]{Wang2001,Kuntz2010,Anderson2016,Bogdan2017}, which is inconsistent with the stellar feedback enrichment scenario.

It is useful to compare our results here with those on the 30 Doradus nebula -- the starburst region in the Large Magellanic Cloud.
 \citet{Cheng2021} have found that the spectrum of the diffuse X-ray emission from the nebula can also be well modeled with the lognormal plasma, although its mean temperature is found to be 0.3~keV, significantly  higher than the upper limit found here for the spectra extracted over the relatively large regions in \sou\ (Table~\ref{t:spec-spiral-arms}). The lower mean temperature of the plasma in the \sou\ may be partly due to the spectral inclusion of multiple plasma components over vast physical regions. Even the spectra of our on-arm regions contain the large-scale  diffuse X-ray emission contributions in the galactic disk and corona, because only the off-galaxy background is subtracted. Therefore, the effective application of the same modeling for \sou\ demonstrates that the lognormal plasma may be used to effectively account for the temperature distribution across large physical regions.
 
 We further explore limitations and uncertainties of the lognormal plasma modeling.  Even under the simple one-zone (single uniform pressure) isobaric plasma assumption, the interpretation of the physical parameters (e.g., $P_{th}$) is not trivial. As shown in \S~\ref{ss:res-spec}, their inferences depend on the validity of the lognormal plasma modeling over a broad temperature range -- an issue that needs to be carefully examined. 
 Theoretically, it remains unclear as to how well a lognormal plasma may describe the spectrum of diffuse X-ray emission from a complex system like a galaxy. Existing studies based on computer simulations are still too limited to allow for a quantitative comparison with our results. Nevertheless, studies such as those conducted by \citet{Vijayan2021} are useful for finding qualitative trends. Their SFR10 simulation of a galaxy  with a  Milky-Way-like mass  and a SFR $= 10 {\rm~M_\odot~yr^{-1}}$, for example, shows that the hot circumgalactic medium exhibits a general correlation of $T \propto 1/n$ over the $10^5$ to $10^7$~K range, as expected in an isobaric state. But around this correlation is also a large dispersion, which may be expected for hot plasma with a range of thermal pressure and is not accounted for in our simple lognormal plasma modeling. It is thus highly desirable to properly model such diversity in the density and temperature distribution, as well as their intrinsic correlation. In addition, the uncertain parameters of $h_{kpc}$ and $f_h$ also need to be characterized before one can effectively use our derived parameters with confidence. 

In any case, the distribution of the temperature down to $\lesssim 0.1$~keV can have strong implications for the radiative cooling of the plasma. We may estimate the {\sl X-ray} cooling time scale of the plasma as $t_{c} = E_{th}/L_{th,x}$.  The parameters in Table~\ref{t:spec-spiral-arms} indicate that $t_c$ is in the range of $(6-24) \times 10^7 (f_h h_{\rm kpc})^{1/2}$~yrs, shorter in the on-arm regions and longer in the outer disk field. This time scale likely represents an upper limit to the {\sl total} cooling time scale of the plasma. In an isobaric state, for example, the time scale is $\propto T^2$.
For plasma at temperatures smaller than 0.1~keV, the radiation is released mostly in the very soft X-ray to extreme UV range and is hence absorbed by cool interstellar gas. Such low-temperature plasma may partly result from mixing layers between hot plasma and cool gas \citep[e.g.,][]{Begelman1990,Fielding2020,Lancaster2021}. Therefore, heat transfer and radiative cooling may, in principle, consume much of the stellar mechanical feedback energy within the galactic disk of \sou.

\subsection{Structure of the X-ray absorbing gas}\label{ss:x-abs}

 According to the results presented in \S~\ref{s:res}, the absorption seems to play a major role in determining the observed morphology, intensity, and spectrum of the diffuse soft X-ray emission.
 The effectiveness of the absorption is intimately related to the actual line-of-sight distribution of the cool ISM relative hot plasma, which is largely unknown. 
 
Dust lanes are visible across much of the galaxy's surface (Fig.~\ref{f:x-3colorHST}), not only at the leading edges of the majestic spiral arms, but also in regions between them, forming ``spurs" or ``feathers". They may represent clouds that rapidly grew, were then differentially compressed in spiral arms and have later been shear-stretched in inter-arm regions \citep{Shetty2006}. Alternatively, such clouds may represent inflow streams under the bar potential and/or outflows from the galactic disk \citep[e/g.,][]{Kannan2021,Farber2021}. In any case, such dust lanes, partly traced by $^{12}$CO emission, are likely interspersed with \ha- and diffuse X-ray-emitting plasma. While $^{12}$CO can often be found right at dust lanes, they tend to be anti-correlated with the diffuse \ha\ and X-ray emissions.
Take the outstanding SWE as an example. It is positionally coincident with a region that shows relatively strong diffuse \ha\ emission and little optical reddening. The X-ray intensity becomes dimmer systematically with increasing reddening or CI in the region toward the east into inner regions of the galaxy; Fig.~\ref{f:x-3colorHST}). This trend is also confirmed in the spiral arm regions, where a linear anti-correlation between the diffuse soft X-ray intensity and the CI is observed (Fig.~\ref{f:x-CI}).
Assuming that the X-ray absorption is foreground, an observed spectrum is $S_\epsilon = S_{\epsilon,0} e^{-N_H \sigma_\epsilon}$, where $S_{\epsilon,0}$ and $\sigma_\epsilon$ are the X-ray emission spectrum and absorption cross-section as the functions of the photon energy ($\epsilon$) and $N_H \propto$ CI. Thus a linear anti-correlation between the X-ray intensity and CI may be expected only when the absorption is small (i.e., $N_H \sigma_\epsilon \ll 0$). However, this condition is apparently inconsistent with the large X-ray intensity variation range that is more than a factor of 2. At least two complications are not accounted for here. Firstly, our measured X-ray intensity is over a broad band so that the X-ray absorption also depends on the spectral shape of the X-ray emission, which remains quite uncertain. Secondly and probably more importantly, the dusty gas is likely inter-mixed with X-ray-emitting plasma or not completely foreground; only the emission beyond the gas is subject to the absorption. The net effect of these complications, though difficult to quantify, should naturally lead to the flattening of the X-ray absorption dependence on the CI: e.g., from the nearly exponential form to the observed linear anti-correlation.

The intermixing scenario of the X-ray emission and absorption is consistent with existing observations. For example, \citet{Liu2013} show that the optical extinction in \sou\ is dominated by dust in a foreground screen, while \citet{Mosenkov2018} estimate that the average face-on optical V-band depth of such dusty screen is $\sim 1$  for a sample of galaxies. The spectral energy distribution modeling of the dust emission suggests that the dusty gas is probably strongly inhomogeneous and clumpy, although its heating mechanism remains largely uncertain and the line-of-sight projection makes it hard to discern the structure. The anti-correlation of the diffuse soft X-ray intensity with the CI, as detected in the present study, indicates that the responsible dusty gas is mostly on the near side of the hot plasma, which is likely located primarily in a thick disk of a characteristic vertical exponential height of $\sim 1$~kpc \citep{Boettcher2017,Nakashima2018,Mosenkov2018,Jiang2019}. This scale should be a factor of a few greater than the scale height of the stellar light. Therefore, the dusty gas should lie mainly away from the galactic plane and likely represent blown-out supershell walls interacting with a relatively slow rotating gaseous coronae of \sou, 
probably similar to so-called intermediate velocity or even high-velocity dusty clouds or complexes  \citep{Marasco2012}. This interaction could lead to the observed lagging halo of  cool gas \citep[e.g.,][]{Peek2009,Rohser2016} and to the condensation of the coronal gas  \citep{Armillotta2016,Vijayan2020,Hobbs2020,Dutta2021,Gronke2021}.  A systematic study of the interaction, confronted by the observed X-ray absorption properties, will thus help to understand the circulation of the cool dusty gas and plasma at galactic disk/halo interfaces \citep[e.g.,][]{Armillotta2016,vandeVoort2021}. 

\subsection{Plasma energetics}
\label{ss:dis-eng}

We have measured the X-ray luminosity of the diffuse thermal plasma, which can be compared with the stellar mechanical energy input in \sou. A useful comparison is the specific luminosity $R_{X,SFR}\approx 5 \times 10^{39} {\rm~erg~s^{-1}/(M_\odot~yr^{-1})}$, as we have estimated from the thermal spectral component in \S~\ref{ss:res-spec}.
This value is consistent with that estimated for \sou\ based on \xmm\ observations \citep{Owen2009}.

To further probe the link between the luminosity of the diffuse soft X-ray emission and the stellar mechanical energy input, we estimate the CC-SN rate from the SFR within $r_c$ of \sou. Using the local background-subtracted \wise\ 22\micron\ image, the integrated SFR is $\sim 1$ 
${\rm M_\odot~yr^{-1}}$ in the \sou\ disk region or a factor of $\sim 2.3$ higher if  the the central ($43^{\prime\prime}$ radius) region is included.
The corresponding energy input rate in the disk  is then 3 $\times 10^{41} {\rm~erg~s^{-1}}$, assuming the conversion of $\sim 1$ SN per 100~yr per 1 ${\rm M_\odot~yr^{-1}}$ \citep{Botticella2012} and $10^{51}$~erg energy per SN. Therefore, the luminosity of the soft X-ray emission within the disk $\sim 1.7  \times 10^{40}  {\rm~erg~s^{-1}}$ (Table~\ref{t:spec-spiral-arms}) implies that the X-ray emission efficiency of the CC-SN mechanical energy feedback is $\sim$ 10\%, which likely represents the lower limit to the bolometric radiation fraction of the energy input. At present, however, the bolometric luminosity or the radiative cooling rate of the plasma cannot be reliably estimated, because of the very uncertain extinction/absorption correction  (\S~\ref{ss:dis-x-spec}).

Dedicated simulations could be helpful in determining the interplay between the stellar feedback and the ISM. Indeed, existing simulations of active star-forming disk galaxies already show that a large fraction ($\gtrsim 0.2$) of the mechanical energy released by SNe could vent into the galactic halo powering a strong galactic wind. This fraction depends on the condition for star cluster-driven superbubbles to break out of galactic disks, which corresponds to a surface SFR threshold \citep[e.g.][]{Fielding2018,Kim2020} $\sim 0.03 {\rm~M_\odot~yr^{-1}~kpc^{-2}}$, equivalent to
$I_{22}= 4.3 {\rm~DN~s^{-1}~arcsec^{-2}}$ (Equation~\ref{e:SFR-I22}).  Figs.~\ref{f:wise-image} and \ref{f:res-2panels} show that this threshold can be readily met in a good fraction of the spiral arms in \sou. The blowout of hot plasma from these most energetic regions can lead to the local reduction of the X-ray enhancement  (due to the EM decrease) or the sub-linear X-ray intensity dependence on the FUV, to the redistribution of chemically-enriched hot plasma (in the downstreams and to larger galactic radii), and to the heating of the galactic corona, responsible for the galaxy-wide diffuse emission as has been studied with the \xmm\ data \citep{Owen2009}.

It remains unclear as to how the mechanical energy input from massive stars is eventually released from a galaxy like \sou. According to \citet{MacLow1988}, about half of the input energy should initially be in hot gas, while the rest is lost in the formation and kinetic energy of the cool supershells around the superbubbles. However, we have shown that the X-ray luminosity from the galactic disk of \sou\ is $\sim 10\%$ of the energy input. Studies of edge-on disk galaxies with various SFRs have further concluded that galactic coronae can account for only a comparable fraction of the energy input  \citep[e.g.,][]{Li2013}.  One  possibility is that superbubbles are effective acceleration sites of cosmic rays \citep[e.g.][]{Lingenfelter2019,Cheng2021}, which can escape into the galactic halo and beyond without much noticeable X-ray radiation. Another possibility is that \sou\ has a strong galactic wind, which drives the bulk of the hot plasma energy out of the immediate vicinity of the galaxy to regions beyond a few tens of kpc. It is also possible that much of the energy in the hot plasma is transformed to cool gas and/or dust (e.g., via turbulent mixing) and is then radiated in other wavelengths (e.g., extreme UV and IR; \citealt[e.g.,][]{Begelman1990,Fielding2020,Lancaster2021}). 
In any case, the X-ray luminosity limits observed in galactic disks and coronae provide fundamental constraints on any theory about the coupling between the stellar feedback and the ISM.

High desirable are new dedicated simulations, including spiral arm dynamics, to confront the results presented here. They need to have sufficiently high spatial and time resolutions to allow for examination of diffuse hot plasma flows and their interplay with other ISM components, addressing such questions as: How does the hot plasma evolve, dynamically, chemically, and thermally? What is its temperature distribution? How is it mixed with X-ray-absorbing cool gas? The simulations will then enable direct confrontation with the data presented here to test our understanding of the astrophysics underlying the coupling between the feedback and the ISM.

\section{Summary and Conclusions}\label{s:sum}

We have conducted a detailed analysis of diffuse soft X-ray emission in the field of \sou, including two outstanding extended background sources. A background FR II radio galaxy  and its radio lobes are found to be associated with part of a linear hard X-ray feature. 
A strongly extended emission peak is attributed to a cluster of galaxies at $z \sim 0.5$ and with an X-ray luminosity of $\sim 3.1 \times 10^{43} {\rm~erg~s^{-1}}$. The X-ray absorption inferred from the spectrum of this cluster is consistent with the estimate based on HI and CO measurements in the galactic disk of \sou. The focus of our study is, however, chiefly on spatial and thermal properties of the diffuse X-ray emission in the  disk, especially in and around its two grand spiral arms, and on gaining insights into the coupling of the stellar feedback with the ISM. Our main results and conclusions are as follows: 

\begin{enumerate} 
\item The enhancement of the diffuse X-ray emission is strongly correlated with the spiral arms,  is bordered by dust lanes on the upstream side of the spiral arms, and is interspersed with $^{12}$CO  emission peaks, as well as \ha\ and FUV emissions. Therefore, the enhancement is closely connected to the feedback from massive stars and may be globally characterized by the ratio $L(0.5-2{\rm~keV})/{\rm SFR} = 5 \times 10^{39} {\rm~erg~s^{-1}/(M_\odot~yr^{-1})}$. 

\item We confirm the square-root dependence of the X-ray enhancement on the FUV intensity, which is first observed in M101.  This sub-linear dependence can be understood as the convolution of the FUV emission evolution of massive stars with the heating, expansion, outflow, and potentially cooling of the diffuse hot plasma.

\item Essentially all relatively bright features or peaks of the FUV emission show enhanced diffuse emission in the 0.45-1~keV band. This enhancement tends to decrease with increasing distance from the galactic center, which can be qualitatively explained by the decreasing metallicity and density/pressure of the ISM toward the outer disk of the galaxy.

\item  We reveal enhanced X-ray emission in the downstreams of the spiral arms. This enhancement tends to fill those so-called inter-arm fork/void regions of the galactic disk and apparently represents the prolonged mechanical feedback from massive stars over a lifetime of $\sim 32$~Myr, coupled with the cooling and/or escaping time of the diffuse hot plasma in the downstreams.

\item The HR ($=S_{0.45-0.7}/S_{0.7-1.0}$) is positively correlated with the diffuse X-ray intensity and generally decreases with the increasing galactic radius in the disk. Part of these trends is due to the soft X-ray absorption, which increases in star-forming regions, especially in the inner disk of the galaxy. But cooling may also play an important role, although the thermal X-ray luminosity of the plasma accounts for only $\sim 10\%$ of the mechanical energy inputs from SNe. 

\item The spectra of the X-ray emission can be reasonably well characterized by an optically-thin thermal plasma with a lognormal temperature  distribution of its emission measure. 
There is no spectral evidence for discrete temperature components. The mean temperature of the lognormal plasma $\bar{T} \lesssim 0.15$~keV indicates that much of its radiation occurs in the photon energy range $\lesssim 0.5$~keV and is not well traced by the X-ray observations. The X-ray spectral shape of the plasma emission can be well characterized by the lognormal temperature dispersion alone, while $\bar{T} = 0.1$~keV is fixed. The metal abundances of the plasma are estimated to be generally super solar, especially in the inner galactic disk region within the co-rotation radius, and tend to be higher than the metallicity measurements based on the integrated UV light of star clusters, although a consistency cannot be completely ruled out because of systematic uncertainties in our X-ray spectral modeling.

\item The differential X-ray absorption inside \sou\ plays an important role in determining both the spectral shape and the intensity distribution of the observed diffuse soft X-ray emission. The effect of the differential absorption on the X-ray spectra can be reasonably well characterized by a lognormal distribution of the absorbing-gas column density. The surface intensity of the emission is overall anti-correlated with the dust reddening in an approximately linear fashion. 

\item The X-ray emission and absorption are most likely inter-mixed globally, which may represent outflows of multi-phase gas from recent massive star-forming regions. Indeed, a good fraction of them, as judged by their surface SFR, seem to be sufficiently energetic to produce such outflows.
The geometry of the inter-mixing remains largely unknown, however, which prevents a reliable estimation of the X-ray absorption effect and hence the radiative cooling rate of the diffuse hot plasma in the galactic disk. 

\item The bulk of the mechanical energy feedback from massive stars is not released in the soft X-ray. Nevertheless, the X-ray  measurements provide fundamental constraints on any theory for the feedback/ISM coupling. Possible additional channels for the feedback energy release include the cosmic-ray, galactic wind, and/or cooling of the hot plasma via mixing with cool gas/dust and radiating at longer wavelengths. It will be important to directly confront the X-ray  measurements with dedicated simulations, including the spiral arm dynamics, and to probe the astrophysics of the coupling between the stellar feedback and the ISM. 
   
\end{enumerate}

\section*{Acknowledgments}

We thank the referee for constructive comments, which helped to improve the paper, appreciate the help of Yingjie Chen in the implementation of the {\small VLNTD} and {\small LNABS} models, and are grateful to Tom Jarrett for sharing the WISE 22\micron\ and radio continuum images of \sou\ used in our multi-wavelength comparison and to Brian Alden for sharing his ClusterPyXT routine, which helped in our construction of the spectral maps.  The research of D.Q. Wang and Y.X. Zeng at University of Massachusetts was partly supported by NASA grants, while \'Akos Bogd\'an acknowledges support from the Smithsonian Institution. 

\section*{DATA  AVAILABILITY}

The X-ray data on \sou\ as described in \S~\ref{s:obs} include all \chandra\ ACIS-S observations taken on the galaxy before 2020 and are available in the \chandra\ data archive  (https://asc.harvard.edu/cda/). Processed data products underlying this article will be shared on reasonable request to the authors.

\bibliographystyle{mnras}
\bibliography{ms.bbl} 

\appendix
\section{Distinct extended background X-ray sources}\label{a:a1}

Although such background sources may be interesting in their own rights, we here probe their properties mainly  to minimize their confusion with the diffuse X-ray emission intrinsic to \sou\ and to explore their potential utility in the study of the galaxy. 

The background cluster of galaxies mentioned in the main text is apparent at R.A. = 13:37:08, Dec. = -29:53:40 (Fig.~\ref{f:x-3color}A), although  no identification in other wavelength bands is yet known \citep{Long2014}.  The spectrum extracted from a radius of 25\farcs7 around the position can be  well fitted with an {\sl APEC} plasma ($\chi^2/{\rm dof} = 562.85/515$). This fit gives a characteristic temperature of $ 2.7_{-0.3}^{+0.5}$~keV and a redshift of $0.56_{-0.01}^{+0.01}$,   as well as a foreground absorption of a column density  $N_H= 2.7_{-0.7}^{+0.7} \times 10^{21} {\rm~cm^{-2}}$. 
 This column density is consistent with the N$_{\rm HI} +2$N$_{\rm H_2}$ value of N$_{HI} \approx 8 \times 10^{20} {\rm~cm^{-2}}$obtained from the 21~cm line observation  \citep{Wall2016} and N$_{H_{2}} \approx 2\times 10^{21} {\rm~cm^{-2}}$   from CO(1-0) emission \citep{Kuno2007}, using the Galactic conversion factor $X_{CO} = N_{H_2}/I_{CO} = 2 \times 10^{20} {\rm~cm^{-2}~[K~km~s^{-1}]^{-1}}$ \citep{Dame2001}. 
This consistency indicates no significant H$_2$  without CO emission along the line of sight.
Adopting the best-fit redshift, we infer a 0.5–2.0 keV  luminosity of $3.1 \times 10^{43}$~\lumcgs, consistent with the X-ray luminosity-temperature scaling relation of galaxy clusters \citep[e.g.,][]{Giles2016}. 

\begin{figure} 
\centerline{
\includegraphics[width=0.5\textwidth]{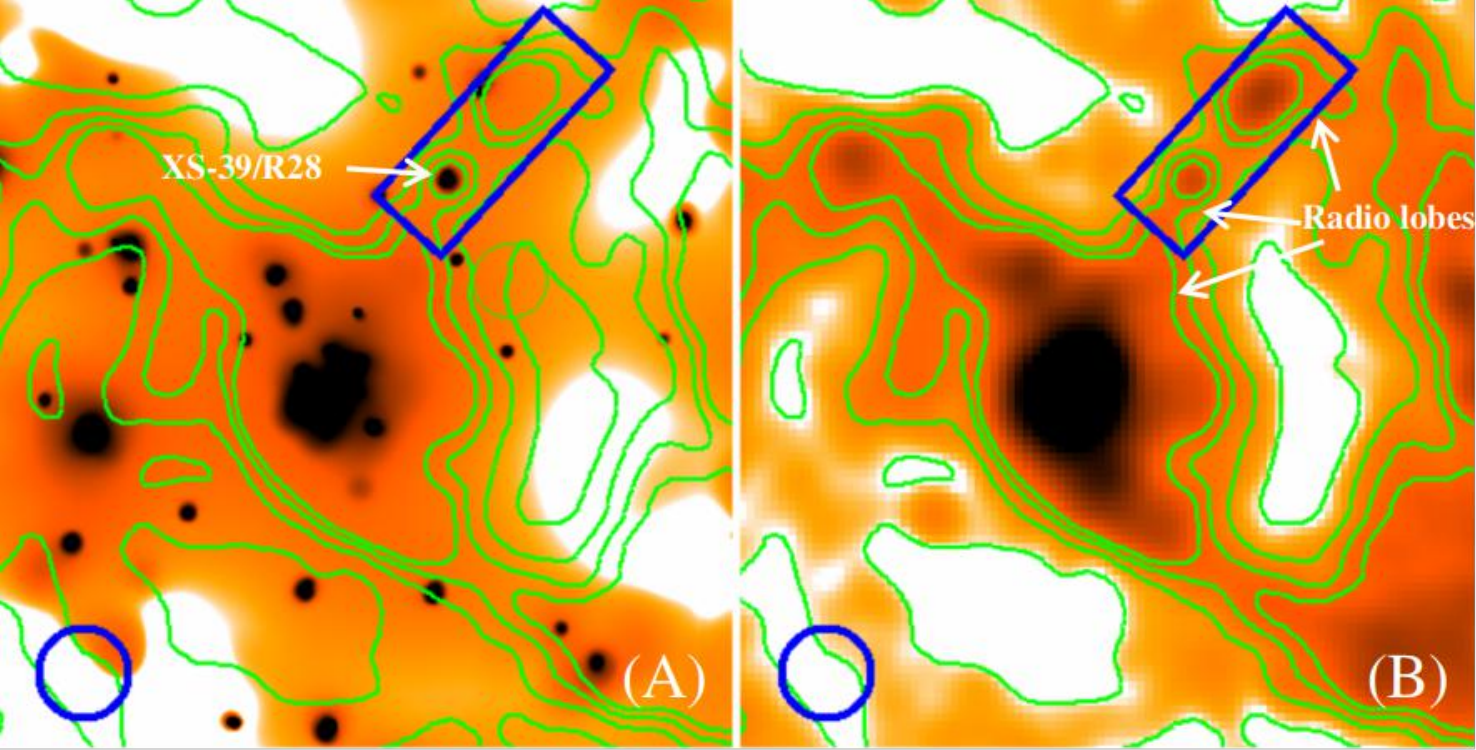}
}\caption{ Close-ups of the \sou\ central field, including the background FR II radio galaxy (enclosed within the blue box):  (A) a Gaussian smoothed \chandra\ intensity image in the 0.45-1~keV band; (B) 6-cm continuum intensity image \citep{Cowan1994}. Overlaid in both panels are the radio intensity contours at 3, 5, 7, and 10$\sigma$ above the background. The lower left blue circle marks a half kpc diameter scale.
}
\label{f:f-NW-x-r}
\end{figure}

On the opposite side of the background cluster relative to the nucleus of \sou\ is a background FR II radio galaxy, which consists of a radio core R-28 and its radio lobes on both sides 
\citep[e.g., Fig.~\ref{f:f-NW-x-r}; ][]{Cowan1994,Maddox2006}. This radio core  spatially coincides with a point-like X-ray source  \citep[XS-39 in][]{Soria2003}. 
Marginally detected at the position of this source is H$\alpha$ line emission with a receding velocity of $\sim 130 {\rm~km~s^{-1}}$, as was  reported by \citet{Dottori2008}. They further claimed the detection of an Fe 
emission line in the \chandra\ spectrum of the source with a small velocity shift corresponding to $z = 0.018$. It was then proposed that the source might be associated with M83 \citep{Dottori2010}. In this scenario, R-28/XS-39 would represent a black hole kicked-off from the nucleus of M83 and could have observational effect on the ISM in the galaxy. Indeed, the scenario has been invoked in the explanation of the pressure gradient observed in molecular gas observed in the region of \sou\ \citep{Wu2015}.

\begin{figure} 
\centerline{
\includegraphics[width=0.5\textwidth]{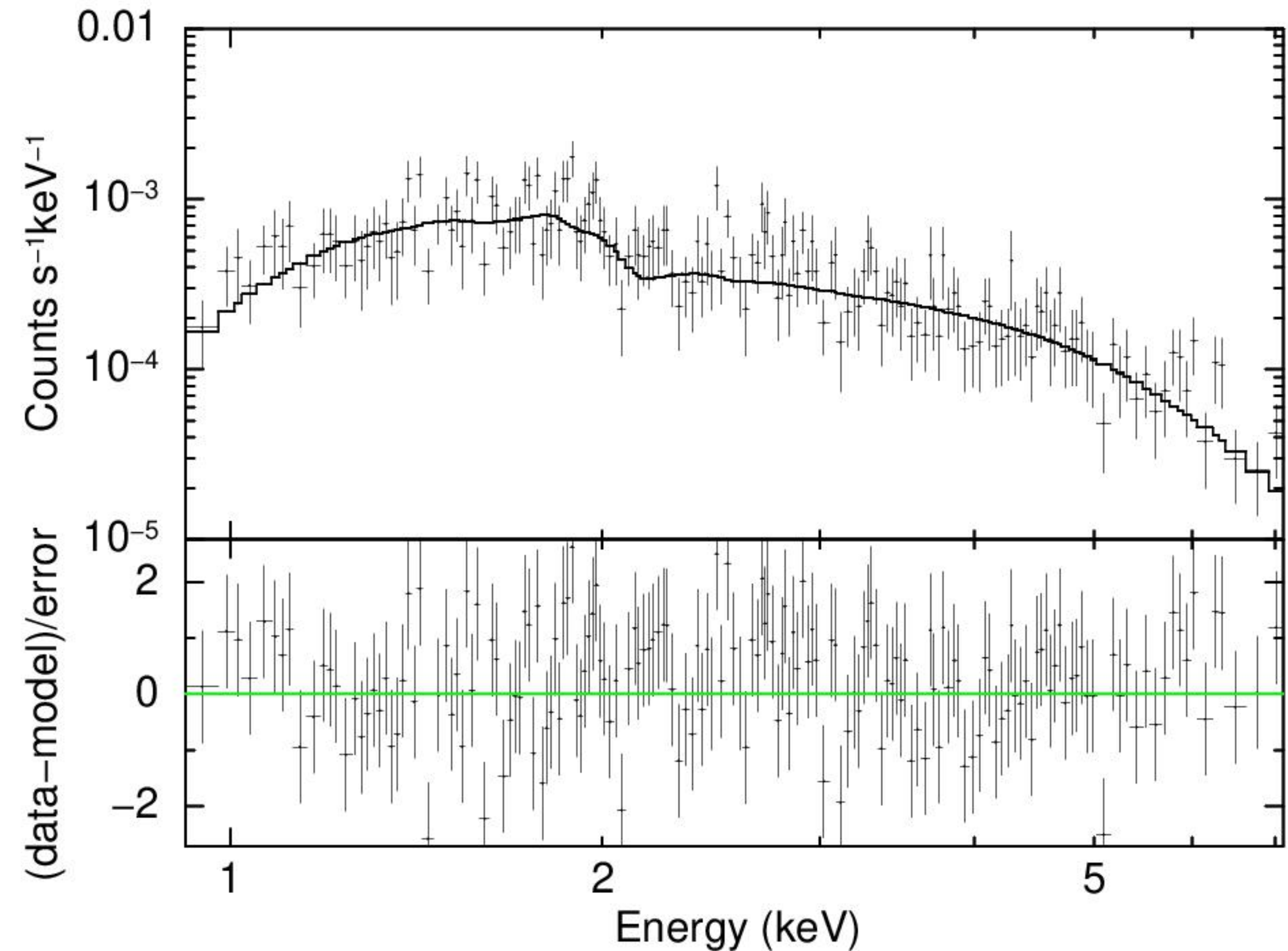}
}\caption{ACIS-S3 spectrum of the background FR II radio galaxy, together with the best-fit power law spectrum.}
\label{f:f-NW-spec}
\end{figure}

We here critically examine the scenario with the deep \chandra\ observations presented here.
The \chandra\ spectrum  used by \citet{Dottori2010} is  from one observation of exposure 51~ks taken in 2000. With a total of only 135 net detected counts, their claimed presence of  the Fe 
line is {\sl assumed} to be intrinsically at 6.7~keV.  Our analysis does not confirm the presence of the line, both statistically and visually (especially on the linear flux density scale, instead of the logarithmic one used in \citet{Dottori2010}).  The lack of a significant Fe K$\alpha$ line in the spectrum of CS-39 is confirmed by our spectral analysis based on the accumulated data from the deep \chandra\ observations 
(e.g., Fig.~\ref{f:f-NW-spec}). We confirm the position coincidence of XS-39 with R-28. The X-ray association with the radio galaxy may even be more extended (e.g., toward to the SE; Fig.~\ref{f:f-spec-maps}).
We speculate that the inverse Compton scattering of radio lobes may be responsible for both the linear morphology and the enhanced diffuse and relatively hard X-ray emission (see Fig.~\ref{f:f-NW-x-r}). 

One may expect a large Faraday rotation of the radio lobes produced by line-of-sight magnetized ionized gas associated with \sou. In principle, by  mapping the Faraday rotation across the radio lobes, one could infer the spatial variation of the line-of-sight magnetic field in the galaxy. 

\section{Hardness ratio calculation}\label{a:HR}
This calculation of the hardness ratio (HR) is based on the X-ray intensities in the two bands (e.g., $S_{0.45-0.7}$ and $S_{0.7-1.0}$. The fluxes are both background-subtracted and exposure-corrected. HR and its error are then calculated as follows:
\begin{equation}
\begin{split}
    S &= \frac{C-B}{E}\\
    \delta S &= \frac{\sqrt{C}}{E}\\
    HR &= \frac{S_{0.7-1.0}}{S_{0.45-0.7}}\\
    \delta HR& = \sqrt{(\frac{\delta S_{0.7-1.0}}{S_{0.45-0.7}})^2+(HR\times\frac{\delta S_{0.45-0.7}}{S_{0.45-0.7}})^2}
\end{split}
\end{equation}
where $C$, $B$ and $E$ are the detected counts, estimated background (including both the non-X-ray (stowed) and local sky X-ray contributions) and exposure (including the energy-dependent effective area of the telescope plus instrument) in each band. The small uncertainty in $B$, which is globally estimated, is neglected in the error propagation. 

\section{Spectral parameter mapping}\label{a:spec_map}

We here first describe the procedure to produce maps of spectral parameters. 
We adaptively determine an extraction circle centered on each pixel to make sure that its spectrum has a signal to noise ratio  $S/N \gtrsim 10$. This is accomplished with the software {\small ClusterPyXT}, which uses the blank sky background to approximately account for the background contribution to the noise \citep{Alden2019}. Our spectral mapping is limited to those pixels with the extraction circle radius less than 50\as.  Each extraction includes the calculation of the local effective area as a function of photon energy to produce the effective area file  ({\sl arf}), while a single spectral matrix file ({\sl rmf}) is adopted, since it varies little across the field. The spectra are then fitted individually with an adopted on-source spectral model to produce the spectral parameter maps.

\begin{figure*} 
\centerline{
\includegraphics[width=1\textwidth]{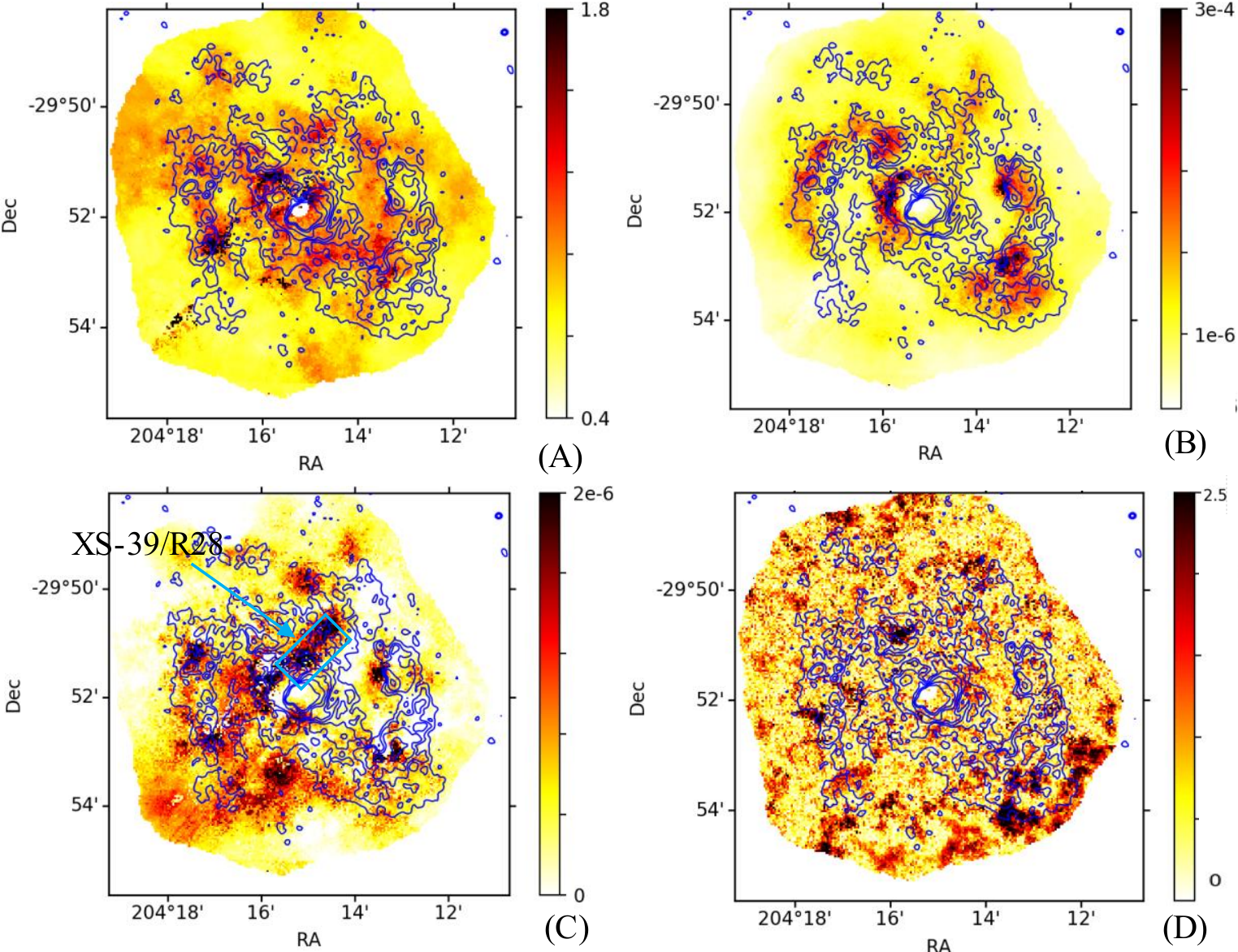}
}
\caption{Maps of the spectral parameters:  (A) the  lognormal temperature dispersion  $\sigma_x$; (B) the normalization of the plasma, scaled to the median spectral extraction area of 0.668 arcmin$^2$; (C) the power law normalization with the region of the background radio galaxy outlined (see also Fig. \ref{f:f-NW-x-r}); and (D) the goodness-of-the-fit approximated as ($\chi^2$-dof)/$\sqrt{2 {\rm dof}}$. The diffuse X-ray intensity contours are the same as those in Fig.~\ref{f:x-dif-gs-UV}.
}
\label{f:f-spec-maps}
\end{figure*}

By adopting the lognormal plasma plus power law model, we map out its spectral parameters  in Fig.~\ref{f:f-spec-maps}. They include the lognormal temperature dispersion,  the normalization of the plasma, the power law normalization, goodness-of-the-fit ($\chi^2$-dof)/$\sqrt{2 {\rm  dof}}$ (under the normal approximation of the $\chi^2$ probability distribution). As in the main text, we fix $\bar{T}$ and $\Gamma$. We further fix the foreground X-ray absorption with a single representative intrinsic absorption column density or its approximate mean value of $N_H = 1 \times 10^{21} {\rm~cm^{-2}}$ (Table~\ref{t:spec-spiral-arms}), plus the Galactic absorption, to minimize the degeneracy that a freely fitting $N_H$ would introduce, especially when the counting statistics of the individual spectra in the  mapping is poor.

The interpretation of these maps is tricky because of the degeneracy among the model parameters. The plasma normalization, for example, is strongly anti-correlated with $\sigma_x$, which partly explains the relatively small normalization at the leading edge of the southeast spiral arm, where the X-ray intensity is high. Another potential cause for this apparent discrepancy is the assumed constant foreground absorption. The actual absorption may be significantly higher in the region southwest of the galactic center, as well as at the leading edges of the spiral arms. Adopting a more realistic absorption, the normalization would be higher. We decide not to increase the modeling complexity in the spectral parameter mapping. Fundamentally, better data are needed to improve the constraints on the spectral properties of the X-ray emission from \sou. 

\end{document}